\documentclass[draftcls, onecolumn]{IEEEtran}

\usepackage{amsfonts}

\usepackage{amssymb}
\usepackage{amsmath}
\usepackage{bbm}
\usepackage{bm}
\usepackage{dsfont}
\usepackage{graphicx}

\usepackage{enumitem}
\usepackage{tikz}
\usepackage{subfigure}

\newcommand{\myin}{{x}}
\newcommand{\mysin}{{x'}}
\newcommand{\myOut}{{\mathcal{Y}}}
\newcommand{\myout}{y}

\newcommand{\myChv}[1]{W_{#1}}
\newcommand{\myCh}[2]{W_{#1}(#2)}
\newcommand{\myChn}[2]{\bm{W}_{#1}(#2)}
\newcommand{\myChnv}[1]{\bm{W}_{#1}}

\newcommand{\braket}[2]{#1^\dagger #2}

\newcommand{\myf}{f}
\newcommand{\myF}{\bm{f}}

\newcommand{\myT}{\mathcal{T}}
\newcommand{\Code}{\bm{\mathcal{C}}}
\newcommand{\myIn}{\mathcal{X}}
\newcommand{\dmin}{d_{\text{min}}}
\newcommand{\dB}{d_{\text{B}}}
\newcommand{\dH}{d_{\text{H}}}
\newcommand{\dL}{d_{\text{L}}}

\DeclareMathOperator{\Tr}{Tr}

\newcommand{\Pe}{\mathsf{P}_{\text{e}}}
\newcommand{\Pem}{\mathsf{P}_{\text{e}|m}}
\newcommand{\Pemax}{\mathsf{P}_{\text{e,max}}}

\newtheorem{example}{Example}
\newtheorem{theorem}{Theorem}
\newtheorem{lemma}{Lemma}
\newtheorem{remark}{Remark}

\begin{document}

\sloppy

\title{Elias Bound for General Distances and Stable Sets in Edge-Weighted Graphs} 

\author{Marco~Dalai,~\IEEEmembership{Member,~IEEE}
\thanks{M. Dalai is with the Department
of Information Engineering, University of Brescia, Italy, e-mail: marco.dalai@unibs.it

Parts of these results were first presented in \cite{dalai-ISIT-2014a}.
}
}

\maketitle

\begin{abstract}
This paper presents an extension of the Elias bound on the minimum distance of codes for discrete alphabets with general, possibly infinite-valued, distances. The bound is obtained by combining a previous extension of the Elias bound, introduced by  Blahut, with an extension of a bound previously introduced by the author which builds upon ideas of Gallager, Lov\'asz and Marton. 
The result can in fact be interpreted as a unification of the Elias bound and of Lov\'asz's bound on graph (or zero-error) capacity, both being recovered as particular cases of the one presented here. 
Previous extensions of the Elias bound by Berlekamp, Blahut and Piret are shown to be included as particular cases of our bound. Applications to the reliability function are then discussed.
\end{abstract}

\section{Introduction}
\label{sec:Intro}
A central problem in coding theory is that of determining the asymptotic performance of optimal codes  when the block length is sent to infinity. Mathematically, this general problem can be formalized in different ways. An example is that of determining the asymptotic minimum distance of binary codes at a given rate. Another example is that of determining the largest possible rate for zero-error communication for a channel over which certain pairs of symbols cannot be confused. Again, another question is the determination of the asymptotic behavior of the probability of error of optimal codes at a given rate. In this paper, we consider a generalization of the first example, that is, the study of the minimum distance of codes at given rate. The proposed approach, however, borrows ideas from bounds originally developed in the context of the second example, that is bounds on the zero-error capacity of channels.

The Elias bound is certainly one of the most famous bounds on the minimum distance of codes. Originally developed for binary codes, it was later extended by Berlekamp \cite{berlekamp-book-1984}, Blahut \cite{blahut-1977} and Piret \cite{piret-1986} to more general contexts and for particular distances. In this paper, we present an extension of the bound to general, possibly infinite-valued, distances. Allowing infinite distances, we handle in a unified way bounds on the minimum distance of codes and bounds on graph (or zero-error) capacity. In fact, we propose a method which builds upon a combination of the Elias bound with the Lov\'asz theta function to bound the minimum distance of codes even in cases where some pairs of symbols have infinite distance. The derived bound will be shown to include as special cases both the mentioned previous extensions of the Elias bound and Lov\'asz' \cite{lovasz-1979} and Marton's \cite{marton-1993} bounds on graph/zero-error capacity.

The bound derived in this paper represents an evolution of some results presented in \cite{dalai-ISIT-2013b}, \cite{dalai-TIT-2013}. The focus was there on the Bhattacharyya distance as a mean for bounding the reliability function of channels. In deriving the new bound, we present the result with generality for an arbitrary distance. We then discuss the particular application to the Bhattacharyya distance and to other distances that can be used to bound the reliability function.

\section{Notation and Problem Definition}
\label{sec:notations}
\subsection{Minimum Distance of Codes}
Let $\mathcal{X}$ be a discrete set and let $d$ be a function $d:\myIn\times\myIn\to\mathbb{R}^+\cup \{+\infty\}$ such that for all $x,x'\in \mathcal{X}$
\begin{align*}
d(x,x') & \geq 0\\
d(x,x')&=d(x',x)\\
d(x,x) & = 0.
\end{align*} 
We will refer to such a function $d$ as a ``distance'', although as seen above we do not really require all the properties of a distance. We stress that $d$ is allowed to take value $+\infty$ for some pairs of symbols, a case which will be of practical interest in this paper.
We extend the distance to sequences of symbols defining, for $\bm{x}=(x_1,\ldots,x_n)$ and $\bm{x}'=(x_1',\ldots,x_n')$,
\begin{equation}
d(\bm{x},\bm{x}'):=\sum_{i=1}^n d(x_i,x_i').
\end{equation}
Note in particular that $d(\bm{x},\bm{x}')=\infty$ iff $d(x_i,x_i')=\infty$ for at least one $i$.
 
A code of length $n$ is a set $\Code=\{\bm{x}_1,\ldots,\bm{x}_M\}$ of elements in $\myIn^n$, that is, sequences of $n$ symbols from $\myIn$ called codewords. The rate of the code is $R=\log M/n$.
For a given code $\Code$, we define its minimum distance as
\begin{equation}
\dmin(\Code) := \min_{\bm{x}, \bm{x}'\in\Code,\,\bm{x}\neq \bm{x}'} d(\bm{x},\bm{x}').
\end{equation}
For a fixed rate $R$ and block length $n$ we define the optimal minimum distance as
\begin{equation}
d(R,n):=\max_{\Code} \dmin(\Code),
\end{equation}
where the maximum is over all codes of length $n$ and rate at least $R$. Finally, for a fixed $R$ we define the asymptotic normalized optimal minimum distance as
\begin{equation}
\delta^*(R):=\limsup_{n\to\infty} \frac{1}{n} d(R,n).
\end{equation}
Similarly, for $\delta\in [0,\infty]$, we also define the maximum rate achievable by codes with minimum distance $\delta^*$ as
\begin{equation}
R^*(\delta):=\sup \{ R:\delta^*(R)\geq \delta\}.
\end{equation}
We stress that $\delta=\infty$ is allowed in our formulation and, consequently, the value $R^*(\infty)$ is also of importance.

In this paper, we are interested in determining upper bounds on $\delta^*(R)$ and $R^*(\delta)$. We are however also interested in bounding the minimum distance of codes with constant composition (or \emph{type}, see \cite{csiszar-korner-book}). Using the notation of \cite{csiszar-korner-book}, this means that there exists a distribution $P$ such that $\bm{x}\in \mathsf{T}_P^n$, $\forall \bm{x}\in\Code$, where $\mathsf{T}_P^n$ is the set of sequences of length $n$ which contain any symbol $x$ exactly $nP(x)$ times. We call $P$ the composition of the codewords.

We thus introduce the basic quantities that we need to consider in this case. In particular, for a valid composition $P$, we define 
\begin{equation}
d(R,n,P):=\max_{\Code} \dmin(\Code),
\end{equation}
where the maximum is over all codes of length $n$, rate at least $R$, and composition $P$. Similarly, for a fixed $R$, we define 
\begin{equation}
\delta^*(R,P):=\limsup_{n\to\infty} \frac{1}{n} d(R,n, P_n).
\end{equation}
with the constraint that the $P_n$ are valid compositions which tend to $P$ as $n\to\infty$, that is
\begin{equation}
\lim_{n\to\infty}P_n(x)=P(x), \quad \forall x.
\end{equation}

\subsection{Matrix/Graph Theoretic Formulation}
\label{sec:graph_formulation}
Given an undirected graph $G$ with vertex set $\mathcal{V}$ and edges $\mathcal{E}\subseteq \mathcal{V}\times\mathcal{V}$, a stable set of nodes is a set of nodes no two (distinct) of which are adjacent in $G$. The stability number $\alpha(G)$ is defined as the size of a largest stable set of nodes. The Shannon capacity of the graph $G$ is defined as
\begin{equation}
C(G):=\lim_{n\to\infty} \frac{1}{n}\log \alpha(G^{\otimes n}).
\label{eq:graphcapacity}
\end{equation}
where $G^{\otimes n}$ is the $n$-fold strong product of the graph $G$ with itself (see \cite{korner-orlitsky-1998} for details).
A constant composition analogue of this quantity can be defined as follows \cite{marton-1993}, \cite{csiszar-korner-1981}. For a composition $P$, consider the subgraph $G^{\otimes n}(P)$ of $G^{\otimes n}$ induced by the subset of the nodes of $G^{\otimes n}$ associated to sequences of composition $P$. We define
\begin{equation}
C(G,P):=\lim_{n\to\infty} \frac{1}{n}\log \alpha(G^{\otimes n}(P_n)),
\label{eq:graphcapacityCC}
\end{equation}
where the $P_n$ are valid compositions which tend to $P$ as $n\to\infty$.

We can extend these definitions so as to include an equivalent formulation of the minimum distance problem.
We consider graphs weighted on the edges, which we identify with the matrix $G$ of the edge weights $g(v,v')$. Here we assume that $g(v,v')\in [0,1]$ and that $g(v,v)=1$.	We say that a set $\mathcal{C}$ of nodes is $\epsilon$-stable if $g(v,v')\leq \epsilon$ for $v,v'\in\mathcal{C}$ with $v\neq v'$. Then we define $\alpha(G;\epsilon)$ as the size of a largest $\epsilon$-stable set. If we now consider the $n$-fold Kronecker power $G^{\otimes n}$ of the matrix $G$, we find that an exponential number of the off-diagonal entries approach zero exponentially fast in $n$. We can then define the $\epsilon$-capacity of the graph as 
\begin{equation}
C(G;\epsilon):=\lim_{n\to\infty} \frac{1}{n}\log \alpha(G^{\otimes n};\epsilon^n).
\end{equation}
Note that, when specialized to $\epsilon=0$, this definition recovers equation \eqref{eq:graphcapacity} if distinct nodes are considered adjacent if and only if $g(x,x')>0$. In this sense, $C(G;\epsilon)$ generalizes $C(G)$ in such a way that $C(G;0)=C(G)$.
Similarly, we can define
\begin{equation}
C(G,P;\epsilon):=\lim_{n\to\infty} \frac{1}{n}\log \alpha(G^{\otimes n}(P_n);\epsilon^n)
\end{equation}
where again the $P_n$ are valid compositions which tend to $P$ as $n\to\infty$.

The problem defined in the previous section can now be reformulated in this setting by considering a graph with vertex set $\mathcal{X}$ and edge weights 
\begin{equation}
g(x,x'):=e^{-d(x,x')},
\label{eq:defgxx'}
\end{equation}
with the convention that $e^{-\infty}=0$.
Since $d(x,x')$ is a measure of the distance between symbols, the function $g(x,x')$ represents a measure of similarity which varies from $1$ to $0$. 
Then, we can observe that we have 
\begin{equation}
R^*(\delta)=C(G,e^{-\delta}).
\end{equation}

We now present two examples which clarify the generality of the considered problem when we allow infinite values for the distance $d(x,x')$ and the usefulness of the graph theoretic formulation.

\begin{example}[Elias Bound for Binary Codes]
\label{ex:binary}
In this case, $\myIn=\{0,1\}$ and $d$ is the Hamming distance $\dH$ defined by setting $\dH(1,0)=1$. Defining the binary entropy function
\begin{equation}
h(t):=-t\log t -(1-t)\log(1-t),
\label{def:binentropy}
\end{equation}
the Elias bound states that if
\begin{equation}
R=\log(2)-h(\lambda), \qquad 0\leq\lambda< 1/2
\end{equation} 
then
\begin{equation}
\delta_{\text{H}}^*(R)\leq 2\lambda(1-\lambda).
\end{equation}
(see the next section for a proof).

The graph representation of the problem is obtained by using the matrix 
\begin{equation}
G=\left(
\begin{array}{cc}
1 & e^{-1}\\
e^{-1} & 1
\end{array}
\right).
\end{equation}
It is easily checked that the $(\bm{x},\bm{x}')$-entry of the matrix $G^{\otimes n}$ has value $e^{- \dH(\bm{x},\bm{x}')}$.
\end{example}

\begin{example}[Lov\'asz' Bound on Graph Capacity \cite{lovasz-1979}]
Lov\'asz' upper bound to the capacity $C(G)$ of an ordinary undirected graph $G$ can be stated as follows. Let $\{u_x\}$ be a set of unit norm vectors in a Hilbert space. We say that $\{u_x\}$ is an orthogonal representation of the graph $G$ if $\braket{u_x}{u_{x'}}=0$ whenever $x$ and $x'$ are not adjacent in $G$. Define the quantity\footnote{We point out that we use a logarithmic definition of the Lov\'asz theta function for ease of comparison with rates and other quantities that we will need in this paper.}
\begin{equation}
\vartheta(G):=\min_{\{u_x\},f}\max_{x}\log\frac{1}{|\braket{u_x}{f}|^2},
\end{equation}
where the minimum is over all orthogonal representations $\{u_x\}$ and unit norm vectors $f$. Then, 
\begin{equation}
C(G)\leq \vartheta(G).
\end{equation}

Note that the problem of determining the graph capacity can be stated in terms of minimum distance of codes by defining a distance $d(x,x')$ such that $d(x,x')=\infty$ if and only if $x$ and $x'$ are not connected in $G$.
Then, $C(G)=R^*(\infty)$.
\end{example}

For the sake of simplicity, we will present our results with a focus on the minimum distance interpretation, but it is clear that an equivalent formulation of each single result can be given in terms of $\epsilon$-capacity of weighted graphs. We will point out some of these formulations when useful.

\section{Previous Extensions of the  Elias Bound}
\label{sec:previousbounds}
Generalizations of the Elias bound to non-binary codes have already appeared in the literature. The main contributions in this direction are those of Berlekamp \cite[Ch. 13]{berlekamp-book-1984}, Blahut \cite{blahut-1977}, and Piret \cite{piret-1986}. 
Those bounds can be considered extensions of the original Elias bound since they are based on the same basic idea.
 For a given code, one first identifies a subset $\myT$ of codewords which are all \emph{packed} in a ball around a properly chosen fixed sequence $\bar{\bm{x}}$. Then, the Plotkin bound is used to bound the minimum distance of the code in terms of the average distance between pairs of distinct codewords in $\myT$ as
\begin{equation}
d_{\text{min}} \leq \frac{1}{|\myT|(|\myT|-1)}\sum_{\bm{x},\bm{x}'\in \myT} d(\bm{x},\bm{x}').
\label{eq:plotkin}
\end{equation}
The average in eq. \eqref{eq:plotkin} can be computed in terms of the componentwise distances as
\begin{align}
\sum_{\bm{x},\bm{x}'\in \myT} d(\bm{x},\bm{x}') & =\sum_{\bm{x},\bm{x}'\in \myT}\sum_{i=1}^n d(x_i,x'_i)\\
& = \sum_{i=1}^n \left(\sum_{\bm{x},\bm{x}'\in \myT}d(x_i,x'_i)\right).
\label{eq:component_plotkin}
\end{align}
Then, the constraints on the compositions of the sequences $\bm{x}, \bm{x}'$ (and possibly $\bar{\bm{x}}$) are used to derive the final bound both in the original formulation for binary codes and in slightly different ways in the more general contexts considered in \cite{berlekamp-book-1984}, \cite{blahut-1977}, \cite{piret-1986}.

In this Section, we first recall for the reader convenience how the approach sketched above is used in the binary case to derive the original bound stated in Example \ref{ex:binary}. Then, we
discuss the extensions proposed in \cite[Ch. 13]{berlekamp-book-1984}, \cite{blahut-1977}, \cite{piret-1986} and provide a high level description of the bound introduce in this work.

\subsection{Binary Codes}
For any $n$-bit sequence $\bm{x}$, let $\mathcal{B}_w(\bm{x})$ be the set of $n$-bit binary sequences which differ from $\bm{x}$ in exactly $w$ positions. It is well known that
that $|\mathcal{B}_w(\bm{x})|=e^{n(h(w/n)+\varepsilon_n)}$, where $h$ is the binary entropy function defined in \eqref{def:binentropy}, and $\varepsilon_n\to 0$ as $n\to \infty$. Hence, by symmetry, any sequence $\bm{x}'\in \{0,1\}^n$ is contained in $\mathcal{B}_w(\bm{x})$ for $e^{n h(w/n)+\varepsilon_n}$ sequences $\bm{x}$. Given any code with $M$ codewords, the total number of codewords counted (with multiplicities) in all the sets $\mathcal{B}_w(\bm{x})$, as $\bm{x}$ runs over $\{0,1\}^n$, is thus $M e^{n h(w/n)+\varepsilon_n}$. Since there are $2^n$ sets $\mathcal{B}_w(\bm{x})$, at least one of them, say $\mathcal{B}_w(\bar{\bm{x}})$, contains $M e^{n (h(w/n)+\varepsilon_n-\log(2))}$ codewords. Note that these codewords are all packed around the sequence $\bar{\bm{x}}$ since they all differ from it in exactly $w$ positions. We can thus use equation \eqref{eq:plotkin} with the choice $\mathcal{T}=\mathcal{B}_w(\bar{\bm{x}}) \cap \Code$. Note that $|\mathcal{T}|=e^{n (R+h(w/n)-\log(2)+\varepsilon_n)}$.
If we call $v_i$ the number of codewords in $\mathcal{T}$ which differ from $\bar{\bm{x}}$ in the coordinate $i$, it can be checked that equation \eqref{eq:component_plotkin} can be continued as
\begin{align}
\sum_{i=1}^n \left(\sum_{\bm{x},\bm{x}'\in \myT}d(x_i,x'_i)\right) & = 
\sum_{i=1}^n  2 v_i (1-v_i).
\label{eq:sum_quadrat}
\end{align}
But, from the definition of $\mathcal{T}$ we must have
\begin{equation}
\sum_{i=1}^n v_i = w |\mathcal{T}|.
\end{equation}
Defining $\lambda_i=v_i/|\mathcal{T}|$, we thus have from \eqref{eq:plotkin} and \eqref{eq:sum_quadrat}
\begin{equation}
d_{\text{min}} \leq \frac{|\mathcal{T}|}{|\myT|-1}\sum_{i=1}^n 2 \lambda_i (1-\lambda_i),
\label{eq:quadrat_binary}
\end{equation}
where the $\lambda_i$ satisfy
\begin{equation}
\frac{1}{n}\sum_{i=1}^n \lambda_i = \frac{w}{n}.
\end{equation} 
Since the function $\lambda(1-\lambda)$ is concave\footnote{This concavity argument is fundamental to all versions of the Elias bound, see Section \ref{sec:squared}.}, we can apply the Jensen inequality to equation \eqref{eq:quadrat_binary} to obtain
\begin{align}
d_{\text{min}}  & \leq 2 n \frac{|\mathcal{T}|}{|\myT|-1} \left(\frac{w}{n}\right) \left(1-\frac{w}{n}\right).
\label{eq:quad_avg_distr}
\end{align}
When $n$ goes to infinity, $|\mathcal{T}|=e^{n (R+h(w/n)-\log(2)+\varepsilon_n)}$ goes to infinity if $w$ is chosen as a function of $n$ in such a way that $w/n$ tends to a limit $\lambda$ such that  $R>\log(2)-h(\lambda)$. We then obtain the bound 
\begin{equation}
\delta^*(R) \leq 2 \lambda(1-\lambda)
\end{equation}
valid for $R>\log(2)-h(\lambda)$. By continuity of $\lambda(1-\lambda)$, we can extend the bound to the case $R=\log(2)-h(\lambda)$.

\subsection{Berlekamp's Bound}
\label{sec:BerlekampsBound}
Berlekamp considers two possible choices of the distance $d$. Assuming $\mathcal{X}$ is the set $\mathbb{Z}_K=\{0,\ldots,K-1\}$, endowed with the usual sum mod $K$, Berlekamp considers the Hamming distance defined by
\begin{equation}
\dH(x,x')=\begin{cases}
0 & \mbox{if } x=x'\\
1 & \mbox{if } x\neq x',
\end{cases}
\end{equation}
and the Lee distance defined by
\begin{equation}
\dL(x,x')=\min(x-x', x'-x).
\end{equation}
Note that in all cases the distance is finite.
The bound in \cite{berlekamp-book-1984} is stated as follows.
Let $V_d^{(n)}$ be the number of sequences of length $n$ with distance at most $d$ from the sequence $\bm{0}=(0,0,\ldots,0)$. Let then 
\begin{equation}
t(R,n)=\frac{1}{n}\min\{d: V_d^{(n)}e^{nR}\geq K^n\}
\end{equation}
and
\begin{equation}
t(R)=\liminf_{n\to \infty} t(R,n).
\end{equation}
Then
\begin{equation}
\delta^*(R)\leq t(R)\left(2-\frac{t(R)}{d(U)}\right),
\end{equation}
where $d(U)$ is the average distance of the points in $\mathcal{X}$ from $0$ under a uniform distribution
\begin{equation}
d(U)=\sum_x \frac{1}{K}d(0,x).
\end{equation}

It is useful to rewrite Berlekamp's bound in a more convenient way. Note that all sequences with a given composition $Q$ are at the same distance from $\bm{0}$, which is given by the expression
\begin{equation}
d(\bm{x},\bm{0})=n\sum_x Q(x)d(0,x).
\end{equation} 
 Since the number of different compositions of sequences in $\mathcal{X}^n$ is polynomial in $n$, the exponential grow of the quantity $V_d^{(n)}$ is determined by the largest composition class at distance at most $d$ from $\bm{0}$. To the first order in the exponent, there are $e^{n H(Q)}$ sequences of composition $Q$, where $H(\cdot)$ is the entropy of a distribution. Hence, we will have
 \begin{equation}
 V_d^{(n)} = e^{n (H(Q)+o(1))},
 \end{equation}
 where $Q$ maximizes $H(Q)$ over the set of compositions satisfying
\begin{equation}
\sum_x Q(x)d(0,x) \leq d.
\end{equation}
Taking the limit as $n\to\infty$, this implies that
\begin{equation}
t(R)=\min_{Q\in \mathcal{Q}(R)} \sum_x Q(x)d(0,x)
\label{eq:t(R)}
\end{equation}
where
\begin{equation}
\mathcal{Q}(R)=\{Q: R+H(Q)\geq \log K\}.
\end{equation}
Since $\mathcal{Q}(R)$ is a convex domain and the objective function is linear, it can be verified with the use of Lagrange multipliers that the optimal $Q^*$ satisfies 
\begin{equation}
Q^*(x)=\eta e^{- \mu d(0,x)}
\end{equation}
where $\eta,\mu$ are parameters chosen so as to satisfy the constraints on $Q$ with equality.

An important remark about this bound is that it relies only on two properties of the considered distances. The first property is that the distances are circularly symmetric, that is 
\begin{equation}
\label{eq:circsymm}
d(x,x')=d(0,x-x')
\end{equation}
and, hence, circular permutations of the elements in $\mathcal{X}$ do not change the cross-distances.
The second property is that the quadratic form
\begin{equation}
\sum_{x,x'}Q(x)Q(x')d(x,x')
\label{eq:quadraticform}
\end{equation}
is a concave function over the simplex of distributions $Q$ (see \cite[Sec. 13.4 and eqs. (13.63)-(13.66)]{berlekamp-book-1984}). 
This requirement is essentially needed in order to apply the Jensen inequality as we did to move from equation \eqref{eq:quadrat_binary} to equation \eqref{eq:quad_avg_distr} in the binary case.
As we will see below, this is a recurrent requirement in all versions of the Elias bound and it is essentially due to the use of the Plotkin bound.

\subsection{Blahut's Bound}
\label{sec:blahut_bound}
Blahut develops his bound on the minimum distance as a tool for bounding the reliability function of discrete memoryless channels. Let $\myCh{\myin}{\myout}$, $x\in \mathcal{X}$, $y\in\mathcal{Y}$, be the transition probabilities of a discrete memoryless channel $W$ with input alphabet 
$\mathcal{X}$ and output alphabet $\mathcal{Y}$.
Blahut considers the Bhattacharyya distance between symbols, defined by
\begin{equation}
\dB(x,x')=-\log \sum_y \sqrt{\myCh{\myin}{\myout}\myCh{\mysin}{\myout}},
\label{def:bhattad}
\end{equation}
under the assumption that $\dB(x,x')$ is finite, which means that the channel $W$ has no zero-error capacity \cite{shannon-1956}. His bound, which applies to a particular subset of channels to be described below, in the most general form states that
\begin{equation}
\delta_{\text{B}}^*(R, P)\leq \min_{V\in\mathcal{V}(P)} \sum_{x,x_1,x_2}P(x)V_x(x_1)V_x(x_2)d(x_1,x_2)
\end{equation}
where $V=\{V_x(x')\}_{x,x'\in\mathcal{X}}$ is a stochastic matrix running over the set
\begin{equation}
 \mathcal{V}(P) =\{V : PV=P,\, I(P,V)\leq R\}.
\end{equation}
Here, $I(P,V)$ is the mutual information with the notation of \cite{csiszar-korner-book}, and $PV=P$ means
\begin{equation}
\sum_{x}P(x)V_x(x')=P(x').
\end{equation}
After optimization of the composition we then have 
\begin{equation}
\label{eq:blahutsbound}
\delta_{\text{B}}^*(R)\leq \max_P \min_{V\in\mathcal{V}(P)} \sum_{x,x_1,x_2}P(x)V_x(x_1)V_x(x_2)d(x_1,x_2).
\end{equation}

The particular class of channels for which this bound is derived by Blahut is that of the so-called non-negative definite channels studied by Jelinek \cite{jelinek-1968}, which are channels for which the matrix $G(\rho)$ with elements (see notations \eqref{eq:defgxx'} and \eqref{def:bhattad})
\begin{equation}
g(x,x')^{1/\rho}=\left(\sum_y \sqrt{\myCh{\myin}{\myout}\myCh{\mysin}{\myout}}\right)^{1/\rho}
\label{eq:g^1/rho}
\end{equation}
is positive semidefinite for all $\rho\geq 1$. 
As we will discuss later (see Lemma \ref{lemma:distances} below), this property implies that the quadratic form in equation \eqref{eq:quadraticform}
 is concave\footnote{Compare also with \cite[Lemma 5]{blahut-1977}; note that there is a sign error in the derivation of the proof and, thus, it is actually proved that the quadratic form is concave and not convex as stated in the lemma.}  in the distribution $Q$, and this is essentially the only property of the distance used in Blahut's derivation. Note in particular that, contrarily to Berlekamp's bound, there is no algebraic structure in the alphabet and no circular symmetry of $\dB(x,x')$ in the sense of \eqref{eq:circsymm},  which is the reason for the more complicated resulting expression in the bound.
 
\subsection{Piret's Bound} 
Piret considers the case where the elements of $\mathcal{X}$ are uniformly spaced points on the unit circle in the plane and the distance is the squared euclidean distance. Note that this is a case where the distance is explicitly assumed to be a circularly symmetric squared euclidean distance.
Piret's bound follows an approach which is similar to Blahut's and can be stated as follows. Let again
\begin{equation}
\mathcal{Q}(R)=\{Q: R+H(Q)\geq \log K\}.
\end{equation}
Then, for any $Q\in\mathcal{Q}(R)$ we have
\begin{equation}
\delta^*(R)\leq \sum_{x,x'} Q(x)Q(x')d(x,x').
\end{equation}
Again, the only properties used by Piret are the circular symmetry of the distance and the concavity of the quadratic form (see \cite[Lemma 4.2]{piret-1986}). 
The latter probably leads Piret to observe that it is difficult to determine the distribution $Q\in\mathcal{Q}(R)$ which gives the best bound, and he thus suggests to search heuristically for a $Q$ which gives good bounds. Interestingly, even if apparently unaware of Berlekamp's approach, he conjectures that the optimal $Q$ is the same $Q^*$ defined in Section \ref{sec:BerlekampsBound}  which minimizes \eqref{eq:t(R)}.

\subsection{Discussion: Squared Euclidean Distances}
\label{sec:squared}
A detailed discussion of the connections between Berlekamp's, Blahut's and Piret's bounds is of interest and, to the best of the author's knowledge, has not been reported previously in the literature. 
We postpone such an analysis to a later section where we show that all three bounds are included as special cases in our bound. It will turn out that our bound includes Blahut's one which in turn
includes Piret's bound, which finally implies Berlekamp's one.

For the moment, instead, we discuss the fundamental common feature that has already emerged in the presentation of the bounds, which is the fact that they all rely on the concavity of the quadratic form \eqref{eq:quadraticform} on the simplex of probability distributions.
In order to better investigate this property, we need the following lemma, whose proof is given in the Appendix.
\begin{lemma}
\label{lemma:distances}
Let $\mathcal{X}$ be a finite alphabet and $d(x,x')$ a distance on $\mathcal{X}$ as defined in Section \ref{sec:notations}. If $d$ is finite, then the following four statements are equivalent
\begin{enumerate}[label={(\alph*)}]
\item\label{item1} The  $|\mathcal{X}|\times |\mathcal{X}|$ matrix $G(\rho)$ with $(x,x')$ element $e^{-d(x,x')/\rho}$ is positive semidefinite for all $\rho> 0$.
\item\label{item2} We have
\begin{equation}
\sum_{x,x'}c(x)c(x')d(x,x')\leq 0, \quad\mbox{provided }\sum_{x\in \mathcal{X}} c(x)=0.
\end{equation}
\item\label{item3} The quadratic form
\begin{equation}
\label{eq:quadratinQ}
\sum_{x,x'}Q(x)Q(x')d(x,x')
\end{equation}
is concave function of $Q$ on the simplex of probability distributions.
\item\label{item4} $(\mathcal{X},d^{1/2})$ is a metric space which can be embedded into a euclidean space, that is, $d$ is a squared euclidean distance.
\end{enumerate}
\end{lemma}

According to the Lemma above, for finite distances the quadratic form \eqref{eq:quadraticform} is concave if and only if the distance $d$ is a squared euclidean distance. That is, there exist points $v_x$ in a euclidean space such that $d(x,x')=\|v_x-v_{x'}\|_2^2$. 
This observation, which does not seem to have been made before in this context,  automatically implies that not only the distance used by Piret, but also those used by Berlekamp and Blahut are squared euclidean distances. Conversely, for the same reason, since the concavity of \eqref{eq:quadraticform} is the only used property, Berlekamp's bound applies to all circularly symmetric squared euclidean distances and Blahut's bound applies to all squared euclidean distances.

We note here that the Hamming distance is trivially representable as a squared euclidean distance in $\mathbb{R}^{|\mathcal{X}|}$ using mutually orthogonal vectors. For the Lee distance, simple embeddings can be found by considering the case of even or odd cardinality separately\footnote{I am indebted to an anonymous correspondent which goes under the nickname ``El Filibustero'' for pointing out these simple embeddings. \label{footnoteFili}}, as shown in Fig. \ref{fig:lee-sqeuclidean}.

For the Bhattacharyya distance, finally, we observe that, as already mentioned, Blahut develops his bound only for the class of non-negative channels studied by Jelinek with the additional assumption that they have no zero error capacity. For these channels, the Bhattacharyya distance is finite and satisfies by definition\footnote{Jelinek only asks the condition for $\rho\geq 1$, but it is not difficult to show that this actually implies it for all $\rho>0$, since the element-wise product of positive semidefinite matrices is positive semidefinite. See also the proof of Lemma \ref{lemma:distances}.} the condition \ref{item1} of Lemma \ref{lemma:distances}. 
It was already observed by Jelinek that the equivalence of conditions \ref{item1} and \ref{item2} in Lemma \ref{lemma:distances}
 was a known fact among algebraists (see his comments to \cite[Th. 2]{jelinek-1968}). However, what was apparently not noticed before in the information theory community, is the equivalence of these conditions with condition \ref{item4} of Lemma \ref{lemma:distances}.
In particular, this implies that for Jelinek's channels the Bhattacharyya distance is a squared euclidean distance. We observe that Jelinek also reports in his Lemma 1 another ``test'' for the condition \ref{item2} in our Lemma \ref{lemma:distances} to be satisfied, namely - using our notation - that the matrix $\tilde{D}$ with elements
\begin{equation}
\tilde{d}(x_1,x_2)=-d(x_1,x_2)+\frac{1}{|\mathcal{X}|}\sum_{x'} d(x_1,x')+\frac{1}{|\mathcal{X}|}\sum_{x'}d(x',x_1)-\frac{1}{|\mathcal{X}|^2}\sum_{x', x''}d(x',x'')
\label{eq:centered_innerprod}
\end{equation}
is positive semidefinite (see the proof of Lemma \ref{lemma:distances} in Appendix). What is curious is that this property is reported by Jelinek but has apparently no importance in his paper, and even if already known in the algebraic community (see \cite{schoenberg-1937}) it was probably not much used outside, at that time. On the contrary, equation \eqref{eq:centered_innerprod}  is by now a fairly well known equation in kernel based learning theory since it represents the condition for a set of points with given cross-distances to be embeddable in a euclidean space. The requirement that the matrix $\tilde{D}$ be positive semidefinite implies that it is the Gram matrix of a set of vectors $\{\tilde{v}_x\}$ and it turns out that the vectors $u_x=v_x/\sqrt{2}$ satisfy $\|u_x-u_x\|^2=d(x,x')$. Hence, Jelinek's channels are precisely those for which the Bhattacharyya distance is a squared euclidean distance. We point out that Jelinek also considers non-negative channels with a zero-error capacity. In this case, the set $\mathcal{X}$ can be partitioned in subsets such that the Bhattacharyya distance is a squared euclidean distance within each subset and it is infinite between symbols from different subsets. So, it can still be interpreted as an euclidean distance if we allow these subsets to be infinitely far apart in the space.

\begin{figure}
\centering

\begin{tabular}{lc}
$|\mathcal{X}|=5:$ & 
\begin{tabular}{rrrrrr}
0 & $\alpha$ & $\alpha$ & 0 & 0\\
0 & $\alpha$ & $\alpha$ & $\alpha$ & 0\\
0 & 0 & $\alpha$ & $\alpha$ & 0\\
0 & 0 & $\alpha$ & $\alpha$ & $\alpha$\\
0 & 0 & 0 & $\alpha$ & $\alpha$\\
\end{tabular}
\\
&\\
$|\mathcal{X}|=6:$ &
\begin{tabular}{rrrrrr}
0 & $1$ & $1$ & $1$ & 0 & 0 \\
0 &  0 & $1$ & $1$ & $1$ & 0 \\
0 & 0 & 0 & $1$ & $1$ & $1$
\end{tabular}
\end{tabular}

\caption{Example of (squared) euclidean embedding of the Lee distance (see footnote \ref{footnoteFili}). The columns of the above matrices, where $\alpha=1/\sqrt{2}$, as points in $\mathbb{R}^5$ and $\mathbb{R}^3$ respectively, have squared euclidean distances which match respectively the Lee distance over $\mathbb{Z}_5$ and $\mathbb{Z}_6$. Extensions to arbitrary values of $|\mathcal{X}|$ is obvious.}
\label{fig:lee-sqeuclidean}
\end{figure}

\subsection{Our Bound}
\label{sec:our_bound}
The extension of the Elias bound that we propose is primarily  motivated by the need to deal with infinite distances. This is a necessary step when considering channels with a zero-error capacity  for which some pairs of symbols cannot be confused. Then, any reasonable distance must take infinite value for non-confusable symbols. This is in fact the case for example with the Bhattacharyya distance.

Our approach is based on a variation of the Plotkin step. In a nutshell, since we want to cope with infinite distances, rather than averaging the pairwise distances $d(\bm{x},\bm{x}')$, we average an exponential function of those distances.
In particular, we use an approach which in a sense corresponds to substituting equation \eqref{eq:plotkin} with 
\begin{equation}
d_{\text{min}} \leq -\rho \log\left( \max_{\bm{x}\in \myT}\frac{1}{(|\myT|-1)}\sum_{\bm{x}'\in \myT\backslash\{\bm{x}\}} e^{-d(\bm{x},\bm{x}')/\rho}\right).
\label{eq:exp_plotkin}
\end{equation}
There is a drawback of course, in that the derivation of the bound must now follow a different route, since it is no longer possible to use eq. \eqref{eq:component_plotkin}. We approach the problem by proposing an extension of the umbrella bound originally introduced in \cite{dalai-ISIT-2013b}. That bound can in fact be interpreted as a variation of the Plotkin bound \eqref{eq:plotkin} in the form of equation \eqref{eq:exp_plotkin}, when there is no constraint on the composition of the codewords $\bm{x}, \bm{x}'$. Here, we propose an extension of the method that allows us to handle composition constraints as is usually done with equation \eqref{eq:component_plotkin}.

\section{Extension of the Elias Bound}

In this section we present our extension of the Elias bound to the case of general, possibly infinite-valued, distances $d$. The extension combines the two basic ideas used in the Elias bound and in Lov\'asz' bound on the zero error capacity. A first step in this unification consists in extending the ordinary Lov\'asz bound to the case of graphs weighted on the edges as presented in Section \ref{sec:graph_formulation}. This was already done in previous works \cite{dalai-ISIT-2013b}, \cite{dalai-TIT-2013}, although the presentation was given for the particular case where the edge weights are related to the Bhattacharyya distances between input symbols of a discrete memoryless channel. We first review that original extension of the $\vartheta$ function in the notation of the present paper, and then we present the additional required extensions and the combination with Elias's bounding procedure.

For the sake of simplicity and for coherence with the literature on Elias' bounds, we develop our procedure with a focus on the rate-distance relation in terms of $\delta^*(R)$ and $R^*(\delta)$ functions. As explained before, however, it is clear that the procedure could be stated solely in terms of weighted graphs and their $\epsilon$-capacities by considering edge weights $g(x,x')$ associated to the given distance $d$ according to $g(x,x')=e^{-d(x,x')}$. We will only briefly mention the analogous expressions in that context and give the main derivation in terms of distances.

\subsection{The $\vartheta(\rho)$ function}

The function $\vartheta(\rho)$ introduced in \cite{dalai-ISIT-2013b} can be defined as follows.
Given the set $\mathcal{X}$ and distance $d$, for a fixed\footnote{Only values of $\rho\geq 1$ were considered in \cite{dalai-ISIT-2013b}, \cite{dalai-TIT-2013}. Here, the way we use $\vartheta(\rho)$ (compare Theorem \ref{th:genplotkin} below with \cite[Th. 1]{dalai-ISIT-2013b}) allows arbitrary positive values of $\rho$.}  $\rho> 0$, an \emph{orthonormal representation of degree $\rho$} of our distance is a set of unit norm vectors $\{u_x\}$ in any Hilbert space such that $|\braket{u_x}{u_{x'}}|\leq e^{-d(x,x')/\rho}$. Call $\Gamma(\rho)$ the non-empty set of all possible such representations
\begin{equation}
\Gamma(\rho) = \left\{ \{u_x\} \, :\,  |\braket{u_x}{u_{x'}}|\leq e^{-d(x,x')/\rho}\right\}, 	\quad \rho> 0.
\label{eq:Gammarho}
\end{equation}
The \emph{value} of an orthonormal representation is the quantity 
\begin{equation}
V(\{u_x\})=\min_{\myf}\max_x\log \frac{1}{|\braket{u_x}{\myf}|^2},
\end{equation}
where the minimum is over all unit norm vectors $\myf$. The optimal choice of the vector $\myf$ is called the \emph{handle} of the representation. The function $\vartheta(\rho)$ is defined as the minimum value over all representations of degree $\rho$, that is,
\begin{align}
\vartheta(\rho) & = \min_{\{u_x\} \in \Gamma(\rho)}V(\{u_x\}).
\end{align}
The result presented in \cite{dalai-ISIT-2013b} can be stated as (a slightly different form of) the following theorem.
\begin{theorem}
For a code $\Code$ of block-length $n$ with $M$ codewords and any $\rho> 0$, we have
\begin{equation*}
\dmin(\Code) \leq - \rho \log \left(\frac{ M e^{-n\vartheta(\rho)} -1}{M-1}\right).
\end{equation*}
\label{th:genplotkin}
\end{theorem}

This result is essentially based on the following Lemma, which we will also need in this paper and that we prove here for convenience.
\begin{lemma}
Let $v_1,\ldots,v_M$ and $w$ be unit norm vectors such that $|\braket{v_i}{w}|^2\geq c>0$ for all $i$. Then 
\[
\max_{i\neq j} |\braket{v_i}{v_j}|\geq \frac{Mc-1}{M-1}.
\]
\label{le:spherical}
\end{lemma}
\begin{IEEEproof}
Let $\Phi$ be a matrix whose $i$-th column is $v_i$. Then, direct computation shows that 
\[
\braket{w}{\Phi}\braket{\Phi}{w}\geq Mc.
\]
Since $w$ is a unit norm vector, $\lambda_{\text{max}}(\Phi \Phi^\dagger)\geq Mc$, where $\lambda_{\text{max}}$ is the largest eigenvalue. This also implies $\lambda_{\text{max}}(\Phi^\dagger\Phi)\geq Mc$. For a matrix $A$ with elements $a(i,j)$, it is known that 
\begin{equation}
\lambda_{\max}(A)\leq \max_{i}\sum_j|a(i,j)|.
\end{equation}
Applying this to $A=\Phi^\dagger \Phi$ we obtain
\begin{align*}
Mc & \leq \lambda_{\text{max}}(\Phi^\dagger\Phi)\\
& \leq \max_{i}\sum_j |\braket{v_i}{v_j}|\\
& \leq 1+(M-1)\max_{i\neq j} |\braket{v_i}{v_j}|
\end{align*}
which implies the statement of the lemma.
\end{IEEEproof}

Given a representation $\{u_x\}$ with handle $f$ achieving $\vartheta(\rho)$, we can associate to a sequence $\bm{x}=(x_1,\ldots,x_n)$ the vector
\begin{equation}
\bm{u}_{\bm{x}}=u_{x_1}\otimes \cdots \otimes u_{x_n}.
\end{equation}
Setting $\bm{f}=f^{\otimes n}$, we find
\begin{align}
|\braket{\bm{u}_{\bm{x}}}{\myF}|^2 & =\prod_{i=1}^n | \braket{u_{x_i}}{\myf} |^2\\ 
& \geq e^{-n\vartheta(\rho)}.
\label{eq:comp_inner}
\end{align}
Hence, for a code $\Code=\{\bm{x}_1,\ldots,\bm{x}_M\}$ Lemma \ref{le:spherical}, used with the vectors $\bm{u}_{\bm{x}_i}$ in place of the $v_i$'s and $\bm{f}$ in place of $w$, implies that 
\begin{equation}
\max_{m\neq m'}|\braket{\bm{u}_{\bm{x}_m}}{\bm{u}_{\bm{x}_{m'}}}|\geq \frac{ M e^{-n\vartheta(\rho)} -1}{M-1}.
\label{eq:maxprod}
\end{equation}
On the other hand, we have 
\begin{align}
|\braket{\bm{u}_{\bm{x}}}{\bm{u}_{\bm{x}'}}| &  = \prod_{i=1}^n |\braket{u_{x_i}}{u_{x_i'}}|\\
& \leq \prod_{i=1}^n e^{-d(x_i,x_i')/\rho}\\
& =e^{-d(\bm{x},\bm{x}')/\rho}
\label{eq:tensorinnerprod}
\end{align}
and, hence,
\begin{equation}
\dmin(\Code)\leq -\rho \log \left( \max_{m\neq m'}|\braket{\bm{u}_{\bm{x}_m}}{\bm{u}_{\bm{x}_{m'}}}|\right).
\label{eq:dminmaxprod}
\end{equation}
Combining equation \eqref{eq:maxprod} and \eqref{eq:dminmaxprod} we obtain Theorem \ref{th:genplotkin}.

When considering the asymptotic regime $n\to\infty$, Theorem \ref{th:genplotkin} implies the following bound on $\delta^*(R)$ (cf. \cite{dalai-ISIT-2013b}, \cite{dalai-TIT-2013}).
\begin{theorem}
\label{th:umbrella}
For any value of $\rho> 0$,
\begin{equation}
\mbox{if } R>\vartheta(\rho), \mbox{ then }\delta^*(R)\leq \rho\vartheta(\rho).
\end{equation}
\end{theorem}

An equivalent formulation of these results can be stated in terms of a weighted graph $G$ by simply letting the graph edge weights $g(x,x')$ play the same role of $e^{-d(x,x')}$ in the definition of the set $\Gamma(\rho)$ in equation \eqref{eq:Gammarho}. Here, we strengthen the notation writing $\vartheta(G,\rho)$ for clarity. Then, the results discussed before can be presented using the definition of $\epsilon$-stable sets as follows.
\begin{theorem}
\label{th:alphaGeps}
For a weighted graph $G$, $\epsilon\in[0,1)$, and any $\rho\geq 0$, we have the bound
\begin{equation}
\alpha(G;\epsilon) \leq \frac{1-\epsilon^{1/\rho}}{e^{-\vartheta(G,\rho)}+\epsilon^{1/\rho}}
\label{eq:eps-stabilitybound1}
\end{equation}
\end{theorem}
Equations \eqref{eq:comp_inner} and \eqref{eq:tensorinnerprod} then essentially imply that $\vartheta(G^{\otimes n},\rho)\leq n \vartheta(G,\rho)$. So,
when used for the graph $G^{\otimes n}$, Theorem \ref{th:alphaGeps} says that 
\begin{equation}
\alpha(G^{\otimes n};\epsilon^n) \leq \frac{1-\epsilon^{n/\rho}}{e^{-n \vartheta(G,\rho)}+\epsilon^{n/\rho}}.
\label{eq:eps-stabilitybound2}
\end{equation}
In the limit of $n \to \infty$, the equivalent of Theorem \ref{th:umbrella} is as follows.
\begin{theorem}
\label{th:C(G,eps)}
For a weighted graph $G$, 
\begin{equation}
\mbox{if } \epsilon<e^{-\rho \vartheta(\rho)} \mbox{ then } C(G;\epsilon)\leq \vartheta(G,\rho).
\end{equation}
\end{theorem}
\begin{remark}
\label{rem:lovaszrecovered}
Note that the standard Lov\'asz bound on the zero-error capacity is obtained by setting $\epsilon=0$, which allows us to use $\rho\to\infty$ thus recovering the bound $C(G)\leq \vartheta(G)$.
\end{remark}
\begin{remark}
In general, with a procedure similar to equations \eqref{eq:comp_inner} and \eqref{eq:tensorinnerprod}, used for moving from equation \eqref{eq:eps-stabilitybound1} to \eqref{eq:eps-stabilitybound2}, we can deduce that for two graphs $G_1$ and $G_2$,
\begin{equation}
\vartheta(G_1\otimes G_2,\rho)\leq \vartheta(G_1,\rho)+ \vartheta(G_2,\rho).
\end{equation}
For the ordinary $\vartheta$ function, we know that the equivalent expression holds with equality \cite{lovasz-1979}. We have not yet investigated whether equality holds also with our extended version, but this will not be needed in the present paper.
\end{remark}

\subsection{Constant Composition Codes}
The first step that we need to consider, for the development of a bound along the Elias scheme, is the extension of Theorem \ref{th:genplotkin} to the case of codes with a constant composition.
Hence, we will first modify our previous approach to bound $\delta^*(R,P)$.
Note that the main property of the function $\vartheta(\rho)$ that we used is the property expressed in equation \eqref{eq:comp_inner}. 
There we really see the reason for the definition of $\vartheta(\rho)$. We built a set of vectors $\{u_{x}\}$ associated to symbols, and an auxiliary vector $\myf$ such that $\myf$ is ``close'' to all possible $u_{x}$. This in turn implies that the vector $\bm{f}$ is close to any vector $\bm{u}_{\bm{x}}$ associated with any sequence $\bm{x}$, no matter what the composition of $\bm{x}$ is.
If we are interested in sequences $\bm{x}$ with a particular composition, however, it can be preferable to pick $\myf$ so that $|\braket{u_x}{\myf}|$ is larger for the symbols $x$ which are used more frequently in the sequence. This leads to a variation of $\vartheta(\rho)$ which is the analogue of the variation of the Lov\'asz theta function introduced by Marton in \cite{marton-1993} (and hence a generalization of the latter).

For a distribution $P$ and for $\rho> 0$, we define
\begin{equation}
\vartheta(\rho,P)=\min_{ \{u_x\} \in \Gamma(\rho), f}\sum_x P(x)\log\frac{1}{|\braket{u_x}{\myf}|^2}.
\end{equation}

With this definition, if $\bm{x}$ is a sequence with composition $P$, and $\{u_x\}$ is a representation with handle $f$ achieving $\vartheta(\rho,P)$, we have
\begin{eqnarray}
|\braket{\bm{u}_{\bm{x}}}{\myF} |^2 & = & \prod_{i=1}^n | \braket{u_{x_i}}{\myf}|^2\\
& =& \prod_{x}|\braket{u_{x}}{\myf}|^{2 n P(x)}\\
& =& e^{n\sum_x P(x)\log |\braket{u_{x}}{\myf}|^2}\\
& = & e^{-n\vartheta(\rho,P)}.
\end{eqnarray}
Consider now a code $\Code$ with $M$ codewords $\bm{x}_1,\ldots,\bm{x}_M$ of composition $P$. If we now apply again Lemma \ref{le:spherical} to the vectors $\bm{u}_{\bm{x}_i}$ we conclude that equation \eqref{eq:maxprod} is simply replaced by
\begin{equation}
\max_{m \neq m'}|\braket{\bm{u}_{\bm{x}_m}}{\bm{u}_{\bm{x}_{m'}}}|\geq \frac{ M e^{-n\vartheta(\rho,P)} -1}{M-1}.
\label{eq:maxprodCC}
\end{equation}
Letting again $n\to\infty$, and using equation \eqref{eq:dminmaxprod}, we have the following result.

\begin{theorem}
For any $\rho>0$,
\begin{equation}
\mbox{if } R>\vartheta(\rho,P), \mbox{ then } \delta^*(R,P)\leq \rho\vartheta(\rho,P).
\end{equation}
\label{th:const_comp_1}
\end{theorem}

\begin{remark}
It is obvious from the definitions that $\vartheta(\rho,P)\leq \vartheta(\rho)$ and, hence, $\max_P \vartheta(\rho,P)\leq \vartheta(\rho)$. This implies that, even after optimization of the distribution $P$, the bound derived here is at least as good as the one that we can derive from Theorem \ref{th:genplotkin}. When $\rho\to\infty$, it can be proved that in fact the equality $\max_P\vartheta(\infty,P)= \vartheta(\infty)$ holds \cite{dalai-winter-2014}. We have not yet investigated if equality holds in general, but this will not be needed in this paper.
\end{remark}

In the graph theory language, this result can be restated as a generalization of Theorem \ref{th:C(G,eps)} to the case of constant composition codes or, using the nomenclature of Marton \cite{marton-1993}, to the case of probabilistic graphs. Note again that, as for Remark \ref{rem:lovaszrecovered}, Marton's result is obtained by setting $\epsilon=0$ and letting $\rho\to\infty$.

\subsection{The Elias Bound}

We now extend further the definition of $\vartheta$ in order to apply the scheme developed by Blahut as a generalization of the Elias bound. What we need now is to extend the definition of $\vartheta(\rho,P)$ to deal with stochastic matrices.
Given a set $\mathcal{A}$, a distribution $F$ on $\mathcal{A}$, and a $|\mathcal{A}|\times |\mathcal{X}|$ stochastic matrix $V=\{V_a(x)\}, a\in\mathcal{A},x\in\mathcal{X}$, we define
\begin{align}
\vartheta(\rho,V|F) & = \sum_a F(a) \vartheta(\rho, V_a)\label{eq:defthetarhoPV}\\
& =\min\sum_{a,x} F(a)V_a(x)\log\frac{1}{|\braket{u_{a,x}}{\myf_a}|^2}\label{eq:defthetarhoPV2}
\end{align}
where the minimum is over all \emph{sequences} of representations $\{u_{a, 1},\ldots,u_{a,|\mathcal{X}|}\}\in\Gamma(\rho)$, $a\in\mathcal{A}$ (one representation for each $a$) and over all sets of unit norm vectors $\{f_a\}$, $a\in\mathcal{A}$ (a different handle for each $a$). 

Consider now the set of optimal representations and optimal handles which achieve $\vartheta(\rho,V|F)$. Let $\bm{a}=(a_1,a_2,\ldots,a_n)$ be a sequence with composition $F$ and define
\begin{equation}
\bm{f}=f_{a_1}\otimes f_{a_2}\cdots\otimes f_{a_n}.
\end{equation}
Assume a sequence $\bm{x}=(x_1,x_2,\ldots,x_n)$ has a conditional composition $V$ given the sequence $\bm{a}$, which means that any symbol $x$ appears in $\bm{x}$ in exactly a fraction $V_a(x)$ of the $nF(a)$ positions in which $a$ appears in $\bm{a}$, for any $a$. Consider the vector
\begin{equation}
\bm{u}_{\bm{x}}=u_{a_1, x_1}\otimes u_{a_2, x_2}\cdots\otimes u_{a_n, x_n}
\end{equation}
Then, we have
\begin{eqnarray}
|\braket{\bm{u}_{\bm{x}}}{\myF} |^2 & = & \prod_{i=1}^n | \braket{u_{a_i, x_i}}{\myf_{a_i}}|^2\\
& =& \prod_{a,x}|\braket{u_{a,x}}{\myf_{a}}|^{2 n F(a)V_a(x)}\\
& =& e^{n\sum_{a,x} F(a)V_a(x)\log|\braket{u_{a,x}}{\myf_a}|^2}\\
& = & e^{-n\vartheta(\rho,V|F)}.
\end{eqnarray}
Applying again Lemma \ref{le:spherical} as we did in our previous bounds, if we have a set of $M$ codewords all with a conditional composition $V$ from a fixed sequence $\bm{a}$ with composition $F$, then 
\begin{equation}
\max_{m \neq m'}|\braket{\bm{u}_{\bm{x}_m}}{\bm{u}_{\bm{x}_{m'}}}|\geq  \frac{M e^{-n\vartheta(\rho,V|F)}-1}{M-1}.
\label{eq:EBD_bound}
\end{equation}

In order to use this inequality for a given code, it is now necessary to consider the possible joint compositions of a subset of codewords with some given fixed auxiliary sequence $\bm{a}\in\mathcal{A}^n$. 
We need the following lemma, where we use the notation 
of \cite{csiszar-korner-book} for types $\mathsf{T}_P^n$ and $V$-shells $\mathsf{T}_V^n(\cdot)$.
\begin{lemma}
\label{le:subcode}
Let $\Code$ be a constant composition code with $\bm{x}\in\mathsf{T}_P^n$, $\forall \bm{x}\in\Code$, and $|\Code|=M$. Let $\hat{V}$ be a conditional composition for sequences on a set $\mathcal{A}^n$ given $\bm{x}\in\mathsf{T}_P^n$ (that is $nP(x)\hat{V}_x(a)$ is an integer) and let $F=P\hat{V}$.
 Then,  there is a subset $\mathcal{T}$ of at least $|\mathcal{T}|=Me^{-n(I(P,\hat{V})+o(1))}$ codewords which all have  joint composition $P\times \hat{V}$ with a fixed sequence $\bm{a}\in \mathsf{T}_{F}^n$.
\end{lemma}

\begin{IEEEproof}
The Lemma is a standard covering argument, and it is essentially a slight generalization of the argument used in \cite[Th. 8]{blahut-1977}. 
It is well known that, for $\bm{x}\in\mathsf{T}_P^n$, $|\mathsf{T}_{\hat{V}}^n(\bm{x})|=e^{n(H(\hat{V}|P)+o(1))}$. On the other hand, if $\bm{a}\in\mathsf{T}_{\hat{V}}^n(\bm{x})$ then $\bm{a}\in \mathsf{T}_F^n$, and $|\mathsf{T}_F^n|=e^{n(H(F)+o(1))}$. Hence, since $|\Code|=M$, at least one sequence $\bm{a}$ is contained in $\mathsf{T}_{\hat{V}}^n(\bm{x})$ for at least $Me^{n(H(\hat{V}|P)-H(F)+o(1))}=Me^{-n(I(P,\hat{V})+o(1))}$ codewords $\bm{x}$.
\end{IEEEproof}

We can now apply the bound of equation \eqref{eq:EBD_bound} for the subset $\mathcal{T}$ of codewords determined in Lemma \ref{le:subcode}. 
Let $V_a(x)=P(x)\hat{V}_x(a)/F(a)$ be the conditional composition of these codewords given the sequence $\bm{a}$. For coherence with our notation, it will be useful to express all quantities in terms of $F$ and $V$ rather than $P$ and $\hat{V}$. From equation  \eqref{eq:EBD_bound} used with the set $\mathcal{T}$ we obtain
\begin{align}
\max_{m \neq m'}|\braket{\bm{u}_{\bm{x}_m}}{\bm{u}_{\bm{x}_{m'}}}| & \geq   \frac{Me^{-n(I(F,V)+\vartheta(\rho,V|F)+o(1))} -1}{Me^{-n(I(F,V)+o(1))}-1}.
\end{align}
Asymptotically as $n\to\infty$, if the rate $R$ is larger than $I(F,V)+\vartheta(\rho,V|F)$, both numerator and denominator in the right hand side of the above equation grow exponentially in $n$ and their ratio is asymptotic to $e^{-n\vartheta(\rho,V|F)}$.
Using again equation \eqref{eq:tensorinnerprod} we have
\begin{equation}
\frac{1}{n}\dmin(\Code) \leq  \rho \vartheta(\rho,V|F) + o(1).
\end{equation}
For fixed $n$, the choice of $F$ and $V$  is  constrained to satisfy the usual type constraints, but asymptotically as $n\to\infty$ these constraints can be neglected.
As a consequence, we have the following theorem.
\begin{theorem}
For given $R$, $P$ and $\rho> 0$, let $F$ be a distribution on a set $\mathcal{A}$ and $V$ be a $|\mathcal{A}|\times |\mathcal{X}|$ stochastic matrix such that $FV=P$. Then,
\begin{equation}
\mbox{if } R > I(F,V)+\vartheta(\rho,V|F),
\quad \mbox{then}\quad\delta^*(R,P)\leq \rho \vartheta(\rho,V|F).
\end{equation}
\label{th:const_comp_2}
\end{theorem}

\begin{remark}
\label{rem:reduction}
We observe that with the choice $\mathcal{A}=\{a\}$ and $V_a(x)=P(x)$ we have $FV=P$, $I(F,V)=0$ and $\vartheta(\rho,V|F)=\vartheta(\rho,P)$. Hence, if $R>\vartheta(\rho,P)$ for a given $\rho$, this particular choice gives the same bound of Theorem \ref{th:const_comp_1}, which is thus included as a particular case in Theorem \ref{th:const_comp_2}.
\end{remark}

In the language of graph theory the result reads as follows.
\begin{theorem}
Under the same conditions of Theorem \ref{th:const_comp_2}, for a weighted Graph $G$ we have
\begin{equation}
\mbox{if } \epsilon < e^{-\rho \vartheta(\rho,V|F)},
\quad \mbox{then}\quad C(G,P;\epsilon)\leq I(F,V)+\vartheta(\rho,V|F).
\end{equation}
\end{theorem}

\section{Analysis of the Bound}

The evaluation of the bound presented in Theorem \ref{th:const_comp_2} is not simple in the general case. A complete theoretical investigation is prevented by the relatively few properties known up to know for the $\vartheta(\rho,V|F)$ function, and even a  numerical study does not seem to be simple in the general case (see Remark \ref{rem:comput_hard} below). In this section, we provide a partial theoretical investigation which is enough to compare our bound with all previous versions of the Elias  bound and with Lov\'asz' and Marton's bound on graph capacity.


\subsection{Binary Channels}
We first give evidence that the proposed bound is a generalization of the Elias bound by showing in detail how the original one for binary channels is recovered as a special case. This shows that, even in the binary case, there is no loss in the use of equation \eqref{eq:exp_plotkin} with the approach based on $\vartheta$ with respect to the standard use of the Plotkin bound \eqref{eq:plotkin} under composition constraints.
In particular, the original bound for binary channels is obtained in the limit $\rho\to\infty$.

Consider a binary alphabet $\mathcal{X}=\{0,1\}$ and distance $d(0,1)=1$. Then, for any $\rho$
it is not difficult to see that one can always take as an optimal representation of degree $\rho$ the two-dimensional vectors 
\begin{align*}
u_0& =[\cos(\alpha), \sin(\alpha)]^\dagger\\
u_1& =[\cos(\alpha), -\sin(\alpha)]^\dagger
\end{align*}
where $\alpha$ satisfies $\cos(2\alpha)=e^{-1/\rho}$. For a given distribution $Q$, let the optimal handle which achieves $\vartheta(\rho,Q)$ be
\begin{align*}
f& =[\cos(\beta), \sin(\beta)]^\dagger.
\end{align*}
Then 
\begin{equation}
\vartheta(\rho,Q)= - 2Q(0)\log \cos(\alpha-\beta)- 2Q(1)\log \cos(\alpha+\beta).
\label{eq:binarytheta}
\end{equation}
where the value of $\beta$ can be determined by minimizing this expression. Upon differentiation and a little of algebra  we find
\begin{equation}
\sin(2\beta)=(Q(0)-Q(1))\sin(2\alpha).
\label{eq:alphabeta}
\end{equation}
The value of $\vartheta(\rho,Q)$ can now be computed analytically by using this relation in \eqref{eq:binarytheta}. The resulting expression is complicated and not 
very useful here. So, we only study the bound of Theorem \ref{th:const_comp_2} asymptotically obtained by letting $\rho\to\infty$ with appropriate choices of $F$ and $V$. We also only study the bound obtained for the uniform composition $P$, since we already know that this is the interesting case for the original Elias bound (see Lemma \ref{lemma:symmetric} below for details).

First note that, for any $V$, $\vartheta(\rho,V|F)\to 0$ as $\rho\to\infty$, which means that we can obtain a bound for any $R$ by choosing $F$ and $V$ such that $I(F,V)<R$.
Let us then choose $\mathcal{A}=\{0,1\}$, $F$ be uniform, and $V$ such that $V_0(1)=V_1(0)=\lambda$, with $\lambda$ such that  $I(F,V)=1-h(\lambda)<R$, where $h(\cdot)$ is the binary entropy function.
If we set $Q=V_0$, then by symmetry we have $\vartheta(\rho,V|F)=\vartheta(\rho,Q)$.
Since  $\cos(2\alpha)=e^{-1/\rho}$, in the limit $\rho\to\infty$ we have $\alpha\to 0$, and from equation \eqref{eq:alphabeta} we deduce that
$\beta\approx \alpha(1-2\lambda)$. The expression for $\vartheta(\rho,Q)$ is then asymptotically
\begin{align*}
\vartheta(\rho,Q) & \approx  - 2(1-\lambda)\log \cos(2\lambda\alpha)- 2\lambda\log \cos(2(1-\lambda)\alpha)\\
& \approx (1-\lambda)(4\lambda^2\alpha^2)+\lambda(4(1-\lambda)^2\alpha^2)\\
& = 4\lambda(1-\lambda)\alpha^2.
\end{align*}
Using again the relation $e^{-1/\rho}=\cos(2\alpha)$ we deduce that
\begin{align}
\rho & = -\frac{1}{\log\cos(2\alpha)}\\
&\approx  \frac{1}{2\alpha^2}.
\end{align}
So, $\rho\vartheta(\rho,Q)\approx 2\lambda(1-\lambda)$. The bound of Theorem \ref{th:const_comp_2} states that for $R>\vartheta(\rho,V|F)+I(F,V)$ we have $\delta^*(R,P)\leq \rho\vartheta(\rho,V|F)$. Since here $\vartheta(\rho,V|F)=\vartheta(\rho,Q)\to 0$ as $\rho\to\infty$, in this limit the theorem says that if $R>1-h(\lambda)$ 
then $\delta^*(R)\leq 2\lambda(1-\lambda)$. This is precisely the Elias bound. One may wonder whether for finite $\rho$ a better bound can be obtained. Unfortunately, a rigorous analysis seems to be painful, but numerical evaluation shows that this is not the case, the optimal bound is  achieved as  $\rho\to\infty$. Analogously, different choices of $\mathcal{A}$, $F$ and $V$ also do not improve the bound.

\subsection{Squared Euclidean Distances}
\label{sec:squared_euclid}
The analysis made above for binary channels can be extended to the general case of any discrete set $\mathcal{X}$ when the distance $d$ is a squared euclidean distance.

Assume then that $d(x,x')$ is a squared euclidean distance. According to Lemma \ref{lemma:distances}, the matrix $G(\rho)$ with entries $g(x,x')^{1/\rho}=e^{-d(x,x')/\rho}$ is positive semidefinite for all $\rho$. Hence, there exist vectors $\{u_x\}$ such that $e^{-d(x,x')/\rho}=\braket{u_x}{u_{x'}}$ for all $x,x'$ and, in particular, these $\{u_x\}$ vectors have unit norm. Hence, the set $\Gamma(\rho)$ defined in equation \eqref{eq:Gammarho} always contains some representations that satisfy all the constraints with equality.
We will consider the bound obtained for one such representation and we will focus in particular on the bound obtained as $\rho\to\infty$.

To make the following derivation easier to follow, we note first that, as $\rho\to\infty$, $\braket{u_x}{u_{x'}}\to 1$ for all $x,x'$, which means that all the vectors tend to concentrate in a very small cap on the unit sphere. Moreover, using the cosine law, as $\rho\to\infty$
\begin{align}
\|u_x-u_{x'}\|^2&=1+1-2\braket{u_x}{u_{x'}}\\
&=2(1-e^{-d(x,x')/\rho})\\
& \approx \frac{2}{\rho} d(x,x').
\end{align}
So, the vectors $u_x$ tend to concentrate on a small cap and they tend to reproduce a scaled version of the original constellation of the given points with their squared distances. We will exploit this fact to show that in the limit $\rho\to\infty$ our bound has a very simple geometric interpretation, which will also allow us to connect our bound to the other ones mentioned before.

For any $F$ and $V$, it is not difficult to see that the optimal choice of the handles $\{f_a\}$ in equation \eqref{eq:defthetarhoPV2} will be such that $\vartheta(\rho,V|F)\to 0$ as $\rho\to \infty$. Hence, in the limit of $\rho\to\infty$, the bound of Theorem \ref{th:const_comp_2} says that if $R>I(F,V)$ then
\begin{equation}
\delta^*(R,P)\leq \lim_{\rho\to\infty} \rho\vartheta(\rho,V|F).
\label{eq:deltarhoinfty}
\end{equation}
So, we are now interested in evaluating the above limit.

For a fixed value of $a$, consider the quantity $\vartheta(\rho,V_a)$ which appears in the definition \eqref{eq:defthetarhoPV}. Let for ease of notation $Q=V_a$, so that we can focus for a moment on the evaluation of  $\vartheta(\rho,Q)$ for a general $Q$ and get rid of $a$.
As mentioned before, we can pick a representation which satisfies $\braket{u_{x}}{u_{x'}}=e^{-d(x,x')/\rho}$, and all these vectors  tend to concentrate in a small cap on the unit sphere as $\rho\to\infty$. The handle $f$ of the representation will surely also be in this small cap and hence $\braket{f}{u_{x}}\to 1$ as $\rho\to \infty$. Let now $\theta_x$ be the angle between the handle $f$ and the vector $u_x$. We have $|\braket{f}{u_{x}}|^2=\cos^2(\theta_x)$ and, since $\theta_x\to 0$, we can use the expansion $-\log(\cos^2(t))= t^2+o(t^2)$, valid for $t\to 0$, to deduce that
\begin{equation}
\log\frac{1}{|\braket{f}{u_{x}}|^2} = \theta_x^2 + o(\theta_x^2).
\label{eq:loccosapprox}
\end{equation}
On the other hand, we have $\|f-u_x\|^2=4\sin^2(\theta_x/2) = \theta_x^2+o(\theta_x^2)$, which implies that 
\begin{equation}
\log\frac{1}{|\braket{f}{u_{x}}|^2}= \|f-u_x\|^2 + o(\|f-u_x\|^2).
\end{equation}
The optimal choice of the handle $f$ will thus be asymptotically such as to minimize a quantity of the form
\begin{equation}
\sum_x Q(x)\left( \|f-u_x\|^2 + o(\|f-u_x\|^2)\right).
\end{equation}
If we neglect for a moment the $o(\cdot)$ term, we notice that the quantity to minimize is precisely the average distortion of a quantizer which uses $f$ for representing the vectors $\{u_x\}$. Without constraints on $f$, it is well known that the choice of $f$ which minimizes the distortion is the centroid and, hence, we expect the handle to satisfy
\begin{equation}
f \approx \sum_{x'} Q(x')u_{x'}.
\label{eq:fapprox}
\end{equation}
However, $f$ must be a unit norm vector and hence we cannot replace the approximation with equality in the above equation\footnote{In practice, as $\rho\to\infty$ all the vectors $u_x$ and $f$ can be considered asymptotically co-planar. However, an accurate estimation of $\vartheta(\rho,Q)$ requires some care in the use of equation \eqref{eq:fapprox}.}.
In order to simplify the discussion, instead of studying the performance obtained for the optimal handle, we show the results obtained for a suboptimal choice, which can however be proved to be the true asymptotically optimal performance with a more detailed analysis.
So, we choose the suboptimal handle
\begin{equation}
f =\frac{\sum_{x'} Q(x')u_{x'}}{\|\sum_{x'} Q(x')u_{x'}\|}.
\label{eq:fsubopt}
\end{equation}
Then, for any $x$ we have
\begin{align}
\theta_x^2\ &  \approx 2(1-\cos(\theta_x))\\
& = 2-2\frac{\sum_{x'} Q(x')\braket{u_{x'}}{u_x}}{\|\sum_{x'} Q(x')u_{x'}\|}.
\end{align}
Using now equation \eqref{eq:loccosapprox}, we have
\begin{align}
\sum_x Q(x)\log\frac{1}{|\braket{f}{u_{x}}|^2}   & \approx 2- 2\frac{\sum_{x',x} Q(x')Q(x)\braket{u_{x'}}{u_x}}{\|\sum_{x'} Q(x')u_{x'}\|}\\
& = 2 - 2 \sqrt{\sum_{x',x} Q(x')Q(x)\braket{u_{x'}}{u_x}}.
\end{align}
The square root in the last expression can be approximated, as $\rho\to\infty$, as follows 
\begin{align}
\sqrt{\sum_{x',x} Q(x')Q(x)\braket{u_{x'}}{u_x}} & = \sqrt{\sum_{x',x} Q(x')Q(x)e^{-d(x,x')/\rho}}\\
& \approx\sqrt{\sum_{x',x} Q(x')Q(x)\left(1-\frac{d(x,x')}{\rho}\right)}\\
& = \sqrt{1- \sum_{x',x} Q(x')Q(x)\frac{d(x,x')}{\rho}}\\
& \approx 1- \frac{1}{2\rho}\sum_{x',x} Q(x')Q(x)d(x,x').
\end{align}
In conclusion, we have the approximation
\begin{equation}
\sum_x Q(x)\log\frac{1}{|\braket{f}{u_{x}}|^2} \approx \frac{1}{\rho} \sum_{x',x} Q(x')Q(x)d(x,x').
\end{equation}
which implies that
\begin{equation}
\lim_{\rho\to\infty} \rho\vartheta(\rho,Q)\leq \sum_{x',x} Q(x')Q(x)d(x,x').
\end{equation} 
If we now use this result for the generic term $\vartheta(\rho,V_a)$ which appears in the definition \eqref{eq:defthetarhoPV}, we find that the right hand side of equation \eqref{eq:deltarhoinfty} can be bounded as
\begin{multline}
\hfill\lim_{\rho\to\infty} \rho\vartheta(\rho,P,V)\leq 
\sum_{a,x_1,x_2} F(a) V_a(x)V_a(x')d(x,x').\hfill
\end{multline} 

So, the bound obtained as $\rho\to\infty$ can be stated as follows.
\begin{theorem}
\label{th:MyEuclidean}
For a squared euclidean distance $d$, for a distribution $F$ on a set $\mathcal{A}$ and a stochastic matrix $V:\mathcal{A}\to\mathcal{X}$ such that $FV=P$ and $R>I(F,V)$, we have the bound
\begin{equation}
\delta^*(R,P) \leq \sum_{a,x_1,x_2} P(x) V_a(x)V_a(x')d(x,x').
\end{equation}
\end{theorem}
To the best of our knowledge, this result is new. When we optimize over $F$ and $V$ to get the best possible bound and over $P$ to get the best possible code, we obtain the following result.
\begin{theorem}
\label{th:genBlahut}
For a squared euclidean distance $d$, we have the bound 
\begin{equation}
\delta^*(R)\leq \max_P \min_{F,V} \sum_{a,x,x'} F(a) V_a(x)V_a(x')d(x,x')
\label{minmaxEliasth}
\end{equation}
where the inner minimum is over the distributions $F$ and $V$ such that $FV=P$ and $I(F,V)\leq R$.
\end{theorem}

\begin{remark}
It can be observed that for the particular choice $\mathcal{A}=\mathcal{X}$ and  $F=P$, the bound takes the form of the Blahut's bound which, as said in Section \ref{sec:squared}, holds for all squared euclidean distances and not just for the Bhattacharyya distance.  
\end{remark}

\begin{remark}
\label{rem:comput_hard}
We observe that the evaluation of \eqref{minmaxEliasth}, as well as Blahut's bound \eqref{eq:blahutsbound}, is more complex than what could seem at first sight because, as we already mentioned in Section \ref{sec:squared}, the objective function in equation \eqref{minmaxEliasth} is concave in $V$ and, hence, the minimization which appears there is not computationally simple.
This problem is essentially the same encountered by Piret in the evaluation of his bound \cite{piret-1986}, and as he suggests, for this type of bounds it may just be preferable to guess good choices of $F$ and $V$ and numerically compute the resulting bound. Since the bound of Theorem \ref{th:const_comp_2} includes the bound of Theorem \ref{th:MyEuclidean}, the same remark applies to it.
\end{remark}

\subsection{Circularly Symmetric Distances}

We now consider the particular case where $\mathcal{X}=\mathbb{Z}_K=\{0,1,\ldots,K-1\}$, endowed with its usual sum, and where the distance $d(x,x')$ is a function of $x-x'$. Note that since $d(x,x')=d(x',x)$ by assumption, $d(x,x')$ is actually a function of $|x-x'|$. Examples of such distances are the Hamming distance, the Lee distance, or the squared euclidean distance for a set of regularly spaced points on the unit circle.
In this case we can simplify our bounds since, due to the symmetry, the uniform composition is optimal for any $R$ as stated in the following Lemma.

\begin{lemma}
\label{lemma:symmetric}
For a circularly symmetric distance, letting $U$ be the uniform distribution, we have
\begin{equation}
\delta^*(R,P)\leq \delta^*(R,U).
\end{equation}
Hence, $\delta^*(R)=\delta^*(R,U)$.
\end{lemma}

\begin{IEEEproof}
The proof of the Lemma is based on a constructive procedure. For a given code $\Code$ of length $n$, with $M=e^{nR}$ codewords of composition $P$ and minimum distance $\dmin(\Code)$, we can construct a code $\tilde{\Code}$ with constant composition $Q$ such that $|Q(x)-1/K|\leq \varepsilon_n$, minimum distance $\dmin(\tilde{\Code})\geq \dmin({\Code})$ and rate $\tilde{R}=R-\alpha_n$, where $\varepsilon_n\to 0$ and $\alpha_n\to 0$ as $n\to\infty$.

Let $\Code=\{\bm{x}_1, \ldots, \bm{x}_M \}$. Let $\bm{X}=(X_1,X_2,\ldots,X_n)$ be a random sequence of uniform independent symbols from $\mathcal{X}$ and set
\begin{equation}
\tilde{\bm{X}}_m=\bm{x}_m+\bm{X}, \quad m=1,2,\ldots,M.
\end{equation}
First note that $d(\tilde{\bm{X}}_m,\tilde{\bm{X}}_{m'})=d(\bm{x}_m,\bm{x}_{m'})$ and, hence, the random code so constructed has the same minimum distance as the original code. Since $\bm{X}$ is uniformly distributed over $\mathcal{X}^n$, $\tilde{\bm{X}}_m$ is also uniformly distributed over $\mathcal{X}^n$. Let $\mathsf{T}(\tilde{\bm{X}}_m)$ be the composition of the sequence $\tilde{\bm{X}}_m$. Let $\mathcal{U}_{\varepsilon}$ be the set of distributions $Q$ such that $|Q(x)-1/K|\leq \varepsilon$. Finally let $\varepsilon_n$ be the smallest $\varepsilon$ for which the following inequality holds
\begin{equation}
\mathsf{P}[\mathsf{T}(\tilde{\bm{X}}_m)\in \mathcal{U}_\varepsilon]\geq 1-\varepsilon.
\label{eq:Palmostuniform}
\end{equation}
Since $\tilde{\bm{X}}_m$ is uniformly distributed over $\mathcal{X}^n$, by the strong law of large numbers, $\varepsilon_n\to 0$ as $n\to\infty$.
Due to equation \eqref{eq:Palmostuniform}, the expected number of codewords $\tilde{\bm{X}}_m$ whose composition is in $\mathcal{U}_{\varepsilon_n}$ is at least $(1-\varepsilon_n)M$. This implies that there exists a sequence $\bar{\bm{x}}\in\mathcal{X}^n$ such that at least $M(1-\varepsilon_n)$ of the codewords
\begin{equation}
\tilde{\bm{x}}_m=\bm{x}_m+\bar{\bm{x}}, \quad m=1,2,\ldots,M,
\end{equation}
have composition in $\mathcal{U}_{\varepsilon_n}$.
Since there are less than $(n+1)^{K}$ different compositions in $\mathcal{U}_{\varepsilon_n}$, at least $M(1-\varepsilon_n)/(n+1)^{K}$ codewords $\tilde{\bm{x}}_m$ have the same composition. This codewords form a constant composition code $\tilde{\Code}$ with rate $\tilde{R}\geq R -O(\log n/n)$, with minimum distance $\dmin(\tilde{\Code})\geq \dmin(\Code)$ and composition in $\mathcal{U}_{\varepsilon_n}$. As $n\to\infty$, this implies that $\delta^*(R,P)\leq\delta^*(R,U)$ and thus that $\delta^*(R)=\delta^*(R,U)$.
\end{IEEEproof}

We can now present the simplification of Theorem \ref{th:const_comp_2} for the case of circularly symmetric distances. Since the uniform composition is always optimal, we can focus on the case where $P$ is the uniform distribution. We can also consider the particular choice $\mathcal{A}=\mathcal{X}$, $F$ uniform on $\mathcal{X}$, and the matrix $V$ to satisfy  $V_x(x')=Q(x'-x)$ for some distribution $Q$. This implies that $I(F,V)=\log K -H(Q)$ and $\vartheta(\rho,V|F)=\vartheta(\rho,Q)$. Then from Theorem \ref{th:const_comp_2}  we deduce the following.
\begin{theorem}
\label{th:cyrcsymmetric_gen}
For a circularly symmetric distance we have the bound
\begin{multline}
\hfill
\mbox{if } R > \log K - H(Q)+\vartheta(\rho,Q),
\quad \mbox{then } \delta^*(R)\leq§ \rho \vartheta(\rho,Q).\hfill
\end{multline}
\end{theorem}

Finally, we can consider the particular case where the distance is a circularly symmetric squared euclidean distance. Then we can combine the simplifications used to obtain Theorems \ref{th:MyEuclidean} and \ref{th:cyrcsymmetric_gen} to obtain the following.
\begin{theorem}
\label{th:PiretgenD}
For a circularly symmetric squared euclidean distance, if $Q$ satisfies $R> \log K- H(Q)$, then 
\begin{equation}
\delta^*(R)\leq \sum_{x,x'} Q(x)Q(x')d(x,x').
\label{eq:quadraticBound}
\end{equation}
\end{theorem}
Note that this bound is essentially the same as given by Piret for the particular case of the squared euclidean distance for regularly spaced points on the unit circle. Hence, Piret's bound is contained as a particular case of Blahut's bound, which is a special case of our own.

We finally show that, Berlekamp's bound can be interpreted as a weakened version of Theorem \ref{th:PiretgenD}.
For a given distribution $Q$, let
\begin{equation}
d(Q)=\sum_x Q(x)d(0,x),
\end{equation}
the average distance from point $0$, that we use as a reference.
Consider again the convex set of distributions
\begin{equation}
\mathcal{Q}(R)=\{Q: R\geq \log K- H(Q)\}.
\end{equation}
We can use in Theorem \ref{th:PiretgenD} any $Q$ in $\mathcal{Q}(R)$, and we choose to use the distribution $Q$ which minimizes $d(Q)$ over $\mathcal{Q}(R)$. Thus, let
\begin{equation}
t:=\min_{Q\in \mathcal{Q}(R)} d(Q)
\label{def:t_min_av_d}
\end{equation}
and let $Q^*$ be a minimizing $Q$.
Then, we have the bound
\begin{align}
\delta^*(R) & \leq \sum_{x,x'} Q^*(x)Q^*(x')d(x,x')\\
 & \leq \max_{Q:\, d(Q)=t\,}\sum_{x,x'} Q(x)Q(x')d(x,x').
\end{align}
Since $d$ is a squared euclidean distance, the quadratic form in the last expression is a concave function and, thus, the maximization can be solved by means of the usual Kuhn-Tucker conditions. It can be observed that this evaluation is the same needed in  Berlekamp's procedure (cf. \cite[eqs. (13.63-13.66)]{berlekamp-book-1984}). The maximizing $Q$ is of the form
\begin{equation}
Q(x)=
\begin{cases} \frac{t}{Kd(U)}+1-\frac{t}{d(U)} &\mbox{if } x = 0 \\ 
\frac{t}{Kd(U)} & \mbox{if } x \neq 0. \end{cases}
\end{equation}
where $U$ is the uniform distribution. For this $Q$ we have
\begin{equation}
\sum_{x,x'} Q(x)Q(x')d(x,x')=t\left(2-\frac{t}{d(U)}\right)
\end{equation}
So, Theorem \ref{th:PiretgenD} implies that 
\begin{equation}
\delta^*(R)\leq t\left(2-\frac{t}{d(U)}\right).
\end{equation}
where $t$ is defined in equation \eqref{def:t_min_av_d}. This is in fact Berlekamp's extension of the Elias bound \cite[Th. 13.67]{berlekamp-book-1984}.
In conclusion, we have shown that our bound includes Blahut's bound as a particular case, which in turns includes Piret's which finally implies Berlekamp's one.

\subsection{Infinite Distances: a Critical Look}
\label{sec:infinite_dist}
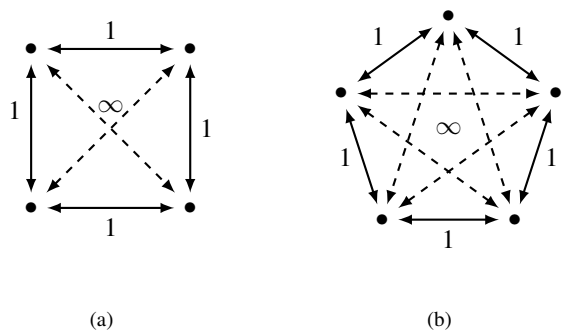
\begin{figure}
\begin{center}
\subfigure[]{
\label{fig:square}
\scalebox{1}{
\begin{tikzpicture}[remember picture]
\begin{scope}
\clip (-2,-2) rectangle (2,2);
{
\foreach \Point/\name in {{(0.75*-1.41, 0.75*1.41)/a}, {(0.75*-1.41, 0.75*-1.41)/b}, {(0.75*1.41, 0.75*-1.41)/c},{(0.75*1.41, 0.75*1.41)/d}}{
   \node (\name) at \Point  {\textbullet}; 
}
}

{
\draw [arrows={latex-latex}, thick] (a)--node[above left]{1}(b);
\draw [dashed, arrows={latex-latex}, thick] (a)--(c);
\draw [arrows={latex-latex}, thick] (a)--node[above]{1}(d);
\draw [arrows={latex-latex}, thick] (b)--node[below]{1}(c);
\draw [dashed, arrows={latex-latex}, thick] (b)--(d);
\draw [arrows={latex-latex}, thick] (c)--node[right]{1}(d);
\node at (0,0.3) {$\infty$};
}
\end{scope}
\end{tikzpicture}}
}
\subfigure[]{
\label{fig:pentagon}
\scalebox{1}{
\begin{tikzpicture}[remember picture]
\begin{scope}
\clip (-2,-2) rectangle (2,2);
{
\foreach \Point/\name in {{(1.5*0.0000, 1.5*1.0000)/a}, {(1.5*-0.9511, 1.5*0.3090)/b}, {(1.5*-0.5878, 1.5*-0.8090)/c},{(1.5*0.5878, 1.5*-0.8090)/d},{(1.5*0.9511,  1.5*0.3090)/e}}{
   \node (\name) at \Point  {\textbullet}; 
}
}

{
\draw [arrows={latex-latex}, thick] (a)--node[above left]{1}(b);
\draw [dashed, arrows={latex-latex}, thick] (a)--(c);
\draw [dashed, arrows={latex-latex}, thick] (a)--(d);
\draw [arrows={latex-latex}, thick] (a)--node[above right]{1}(e);
\draw [arrows={latex-latex}, thick] (b)--node[left]{1}(c);
\draw [dashed, arrows={latex-latex}, thick] (b)--(d);
\draw [dashed, arrows={latex-latex}, thick] (b)--(e);
\draw [arrows={latex-latex}, thick] (c)--node[below]{1}(d);
\draw [dashed, arrows={latex-latex}, thick] (c)--(e);
\draw [arrows={latex-latex}, thick] (d)--node[right]{1}(e);
\node at (0,0) {$\infty$};

}

\end{scope}
\end{tikzpicture}}
}
\end{center}
\caption{The two examples of distances discussed in Section \ref{sec:infinite_dist}.}
\label{fig:polygons}
\end{figure}

In the case of infinite distances Theorem \ref{th:const_comp_2} exhibits both interesting properties as well as clear weaknesses. In this case, we know that even bounds on $R^*(\infty,P)$ and $R^*(\infty)$ (or on $C(G,P)$ and $C(G)$, in the graph theory language) are hard, and we 
first remind that, as mentioned in Remark  \ref{rem:reduction}, Lov\'asz' and Marton's bounds on the capacity of graphs are recovered from Theorem \ref{th:const_comp_2} with a trivial choice of $\mathcal{A}$, $F$ and $V$.
The performance of the bound for $R>R(\infty)$, however, strongly  depends on the particular type of graph $G$ which is induced by finite values of $d$ on $\mathcal{X}\times \mathcal{X}$, and it is certainly not yet satisfactory in the general case.
A general analysis is prohibitively complex, but it will be useful  to consider two particular cases for which we can perform simple sanity checks on our bound (see Figure \ref{fig:polygons}).

Let $\mathcal{X}=\{0,1,2,3\}$ be the vertices of a square and let $d(x,x')=1$ if $x,x'$ are adjacent, while $d(x,x')=\infty$ otherwise (see Figure \ref{fig:square}). The graph induced by finite distances is the square itself and its capacity is $\log(2)$. Due to symmetry, we test the simplified version of the bound given in Theorem \ref{th:cyrcsymmetric_gen}.
If we choose 
\begin{equation}
Q(x)=
\begin{cases}
1-\lambda & \mbox{if } x=0 \\
\lambda & \mbox{if } x=1\\
0 & \mbox{otherwise}
\end{cases}
\end{equation}
and consider the result obtained as $\rho\to\infty$, the bound reduces to the statement that for $R> \log(4)-h(\lambda)$, $\delta^*(R)\leq 2 \lambda(1-\lambda)$. Note that the bound on $\delta^*(R)$ is smaller than $1/2$ at all rates at which it is bounded, that is for $R>\log(2)$ (the capacity of the graph, indeed).
In fact, it can be observed that the bound is exactly the standard Elias bound for binary codes shifted by a quantity $\log(2)$ on the $R$ axis. 
This is in accordance with intuition, since it is not difficult to see that at rates $R>\log(2)$ there are at least $e^{n(R-\log(2))}$ codewords which are all at finite distance and which can be mapped to a binary alphabet without modifying the distances among them. Thus, bounds on $\delta^*(R)$  for the original setting can be deduced from bounds on $\delta^*(R-\log(2))$ for binary codes. This is automatically taken care of in our bound and, 
hence, in this case we can say that the bound is a satisfactory extension of the standard bound. It is not difficult to see that this happens for all even cycles.

Consider instead the case of the pentagon with vertex set $\mathcal{X}=\{0,1,2,3,4\}$, and where we let $d(x,x')=1$ if $x,x'$ are adjacent in the pentagon, while $d(x,x')=\infty$ otherwise (see Figure \ref{fig:pentagon}). The graph induced by finite distances is the pentagon itself and its capacity is $\log(5)/2$ \cite{lovasz-1979}.
If we apply Theorem \ref{th:cyrcsymmetric_gen} with the same choice of $Q$ mentioned above and letting $\rho\to\infty$, we get the bound $\delta^*(R)\leq 2 \lambda(1-\lambda)$ for $R>\log(5)-h(\lambda)$. So, this choice of $Q$ only gives a finite (and reasonably good) bound for $R>\log(5/2)>\log(5)/2$. This was to be expected, since we are essentially not using the Lov\'asz theta function of the pentagon\footnote{We are using a $Q$ which is good enough to obtain the fractional clique covering number of the graph. Actually, there is even no need to use $\vartheta$ functions for this choice of $Q$, since we reduce the problem to binary alphabets. We do not go into these details and leave further analysis to future works.} with this choice of $Q$.
So, finite bounds on $\delta^*(R)$ for $\log(5)/2<R<\log(5/2)$ require, as is obvious, other choices of $Q$ and $\rho$.
A detailed analysis is complicated and, as mentioned, the evaluation of the bound is not simple in general and will be hopefully investigated in more detail in a future work. We can here at least mention that, to the best of our understanding, for rates slightly larger than $\log(5)/2$, no choices of $\rho$ and $Q$ lead to a bound on $\delta^*(R)$ which is as good as the trivial bound $\delta^*(R)\leq1$. 
This is of course a frustrating negative point on our bound. One may ask whether other choices of $\mathcal{A}$, $F$ and $V$ in the original bound of Theorem \ref{th:const_comp_2} would give better results than Theorem \ref{th:cyrcsymmetric_gen}. We tend to exclude this, though we do not have a rigorous prove.

The pentagonal example discussed above shows that, even if the bound has the reasonably good property of including all previous versions of the Elias bound as well as Lov\'asz' and Marton's bounds on graph capacities, it is still  surely not a satisfactory bound in the case of general possibly infinite valued distances. We propose the study of bounds on $\delta^*(R)$ for the pentagonal example proposed here as an interesting open problem which deserves further attention.

\section{Reliability Function}

We present here an important case of application of the bound, that is, its use in bounding the reliability function of classical and classical-quantum channels. We describe this two cases separately for the reader's convenience. 

\subsection{Classical Channels}

Let $\myIn=\{1,2,\ldots,|\myIn |\}$ and $\myOut=\{1,2,\ldots,|\myOut|\}$ be the input and output alphabets of a discrete memoryless channel with transition probabilities 
 $\myCh{\myin}{\myout}$, $\myin\in\myIn,\myout\in\myOut$.
If $\bm{\myin}=(\myin_1,\myin_2,\ldots,\myin_n)$ is a sequence of $n$ input symbols and correspondingly  $\bm{\myout}=(\myout_1,\myout_2,\ldots,\myout_n)$ is a sequence of output symbols, then the probability of observing $\bm{\myout}$ at the output of the channel given input $\bm{\myin}$ is 
\begin{equation*}
\myChn{\bm{\myin}}{\bm{\myout}}=\prod_{i=1}^n \myCh{\myin_i}{\myout_i}.
\end{equation*}
An $(M,n)$ code is a set of $M$ $n$-symbol sequences $\{\bm{x}_1,\ldots,\bm{x}_M\}$, $\bm{x}_m\in \mathcal{X}^n$ associated to $M$ messages $\{1,\ldots,M\}$, and a decoder is a map from the set of output sequences $\mathcal{Y}^n$ to $\{1,2,\ldots,M\}$.
Let $\bm{\myOut}_m$ be the set of output sequences that are mapped to the message $m$ by the decoder. When message $m$ is sent, the probability of error is
\begin{equation*}
\Pem=1-\sum_{\bm{\myout}\in \bm{\myOut}_m} \myChn{\bm{\myin}_m}{\bm{\myout}}.
\end{equation*}
The maximum error probability of the code is defined as the largest $\Pem$, that is,
\begin{equation*}
\Pemax=\max_{m}\Pem.
\end{equation*}
Let $\Pemax^{(n)}(R)$ be the smallest maximum error probability among all codes of length $n$ and rate at least $R$. The reliability function is defined as (see \cite{shannon-gallager-berlekamp-1967-1} for more details)
\begin{equation}
E(R)=\limsup_{n\to\infty} -\frac{1}{n}\log P_{e,\max}^{(n)}(R).
\label{eq:E(R)_def_class}
\end{equation}

\begin{figure}
\centering
\includegraphics[scale=0.8]{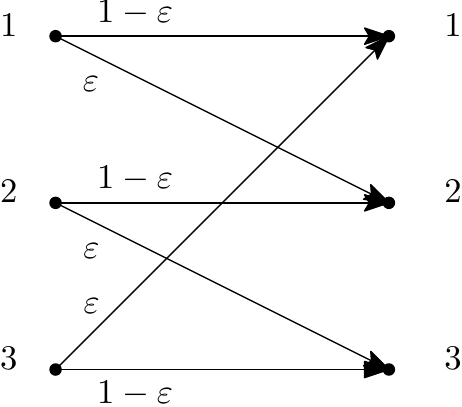}
\caption{The ``ternary unilateral channel'' of \cite[Fig. 3]{shannon-gallager-berlekamp-1967-2}.}
\label{fig:ternarychannel}
\end{figure}

For any channel, for a given code, the probability of error $\Pemax$ is lower bounded by the probability of error in any binary hypothesis test between two codewords. In a binary hypothesis test between codewords $m$ and $m'$, an extension of the Chernoff Bound allows to assert that the minimum error probability $\Pe$ vanishes exponentially fast in the block length $n$ and that  \cite{shannon-gallager-berlekamp-1967-2}
\begin{equation*}
\log\frac{1}{\Pe}\leq D_{\text{C}}(\myChnv{\bm{\myin}_m},\myChnv{\bm{\myin}_{m'}})+o(n).
\end{equation*}
where $D_C(\cdot, \cdot)$ is the Chernoff distance between two distributions defined by
\begin{equation}
D_{\text{C}}(Q_1,Q_2)=-\log\inf_{0< s< 1} \sum_\myout Q_1(y)^{1-s}Q_2(y)^s.
\end{equation}
Note that we use a different notation for the Chernoff distance because it is not additive, in the sense that, in general,
\begin{equation}
D_{\text{C}}(\myChnv{\bm{\myin}_m},\myChnv{\bm{\myin}_{m'}}) \neq \sum_{i=1}^n D_{\text{C}}(\myChv{\myin_{m,i}},\myChv	{\myin_{m',i}}).
\label{eq:Chernonadditive}
\end{equation}
Using the above considerations, $E(R)$ can be bounded as
\begin{equation}
E(R)\leq \frac{1}{n}\min_{m\neq m'} D_{\text{C}}(\myChnv{\bm{\myin}_m},\myChnv{\bm{\myin}_{m'}})+o(1).
\label{eq:ERminchernoff}
\end{equation}
Hence, upper bounds on $E(R)$ can be deduced by determining upper bounds on  the minimum Chernoff distance which appears on the right hand side of \eqref{eq:ERminchernoff}. Due to equation \eqref{eq:Chernonadditive}, we cannot apply our bound on the minimum distance directly to the Chernoff distance, but we can use additive distances which upper bound it. The Bhattacharyya distance can be used for this purpose; it can be proved that 
\begin{equation}
\dB(x,x')\leq D_C(W_x,W_{x'})\leq 2 \dB(x,x')
\label{eq:dbchernoffineq}
\end{equation}
For the so called \emph{pairwise reversible channels} \cite{shannon-gallager-berlekamp-1967-2}, we have equality on the left hand side for all $x,x'$ and hence
\begin{equation}
E(R)\leq \delta_{\text{B}}^*(R).
\end{equation}
Thus, our bounds on $\delta_{\text{B}}^*(R)$ apply directly to $E(R)$. However, for other channels, equality holds on the right hand side of \eqref{eq:dbchernoffineq}. For these channels, the best that we can do in bounding $E(R)$ using $\dB$ is using the inequality
\begin{equation}
E(R)\leq 2 \delta_{\text{B}}^*(R).
\label{eq:ER2db}
\end{equation}
We observe in particular that Blahut's proof of his upper bound on $E(R)$ in \cite[Th. 12]{blahut-1977} only holds for pairwise reversible channels. The problem for general channels comes from \cite[Sec. VI, page 669, second column]{blahut-1977} where it is stated that \emph{``Now $x_{m'}$ and $x_m$ have the same composition [...] ; hence, the first term is zero''}. This statement is not correct, since it would essentially imply that the Chernoff distance between two codewords with the same composition equals their Bhattacharyya distance, which is not always the case. More specifically, in our notation, the quoted statement is that if $\bm{x}$ and $\bm{x'}$ are two codewords with the same composition, then, setting
\begin{equation}
\bm{Q}(\bm{y})=\frac{\sqrt{\bm{W}_{\bm{x}}(\bm{y})\bm{W}_{\bm{x}'}(\bm{y})}}{\sum_{\bm{y}'}\sqrt{\bm{W}_{\bm{x}}(\bm{y}')\bm{W}_{\bm{x'}}(\bm{y}')}}
\end{equation}
we have
\begin{equation}
\sum_{\bm{y}}\bm{Q}(\bm{y})\log\frac{\bm{W}_{\bm{x}}(\bm{y})}{\bm{W}_{\bm{x}'}(\bm{y})}=0.
\end{equation}
This is not true, as proved by the ``ternary unilateral channel'' mentioned in \cite{shannon-gallager-berlekamp-1967-2} and shown in Fig. \ref{fig:ternarychannel}. For the codewords $\bm{x}=(1,2,3)$ and $\bm{x}'=(2,3,1)$, which have the same composition, we have
\begin{equation}
\bm{Q}(\bm{y})=\begin{cases}
1 & \mbox{if } \bm{y}=(2,3,1)\\
0 & \mbox{otherwise}
\end{cases}
\end{equation}
and consequently
\begin{equation}
\sum_{\bm{y}}\bm{Q}(\bm{y})\log\frac{\bm{W}_{\bm{x}}(\bm{y})}{\bm{W}_{\bm{x}'}(\bm{y})}=3\log \frac{1-\varepsilon}{\varepsilon}.
\end{equation} 
One of the consequences of this observation is that, contrarily to what stated in \cite[Sec. I]{blahut-1977}, Blahut's  bound does not include the zero-rate bound of \cite{shannon-gallager-berlekamp-1967-2} for channels that are not pairwise reversible. So, the only proof that the expurgated bound is tight at $R=0$ for any channel without a zero-error capacity remains Berlekamp's complicated proof \cite{berlekamp-thesis}, \cite{shannon-gallager-berlekamp-1967-2}.

For channels that are not pairwise reversible but for which $D_C(W_x,W_{x'})$ is only slightly larger than $\dB(x,x')$, using equation \eqref{eq:ER2db} can be very suboptimal. A possible alternative approach is to introduce an additive variation of the Chernoff distance. We can define
\begin{equation}
{d}_{\tilde{\text{C}}}(x,x')=D_C(W_x,W_{x'})
\end{equation}
and extend $d_{\tilde{\text{C}}}$ additively to sequences. It is then not difficult to see that for sequences $\bm{x}$, $\bm{x}'$
\begin{equation}
{d}_{\tilde{\text{C}}}(\bm{x},\bm{x}')\geq D_{\text{C}}(\myChnv{\bm{\myin}_m},\myChnv{\bm{\myin}_{m'}}).
\end{equation}
So, we have
\begin{equation}
E(R)\leq \delta_{\tilde{\text{C}}}^*(R)
\end{equation}
and we can thus bound $E(R)$ by using our bound for the distance $d_{\tilde{\text{C}}}$. Note in particular that this recovers the particular case of pairwise reversible channels, since $d_{\tilde{\text{C}}}=\dB$ in that case.
On the other hand, for the channel of Fig. \ref{fig:ternarychannel}, as $\varepsilon\to 0$ $d_{\tilde{\text{C}}}(x,x')/\dB(x,x')\to 2$, and we thus recover equation \eqref{eq:ER2db} which, as $R\to 0$, is loose by a factor of two (cf. \cite{shannon-gallager-berlekamp-1967-2}).

\subsection{Classical-Quantum Channels}
Consider a classical-quantum channel with input alphabet $\myIn=\{1,\ldots,|\myIn|\}$ and associated density operators $S_\myin$, $\myin\in\myIn$, in a finite dimensional Hilbert space $\mathcal{H}$. The $n$-fold product channel acts in the tensor product space $\bm{\mathcal{H}}=\mathcal{H}^{\otimes n}$ of $n$ copies of $\mathcal{H}$. To a sequence $\bm{\myin}=(\myin_1,\myin_2,\ldots,\myin_n)$ is associated the signal state $\bm{S}_{\bm{\myin}}=S_{\myin_1}\otimes S_{\myin_2}\cdots\otimes S_{\myin_n}$. 
As in the classical case, an $(M,n)$ code is a set of $M$ $n$-symbol sequences $\{\bm{x}_1,\ldots,\bm{x}_M\}$, $\bm{x}_m\in \mathcal{X}^n$ associated to $M$ messages $\{1,\ldots,M\}$.
A quantum decision scheme for such a code is a so-called POVM (see for example \cite{wilde-2013}), that is, a  collection of $M$ positive operators $\{\Pi_1,\Pi_2,\ldots,\Pi_M\}$ such that $\sum \Pi_m \leq \mathds{1}$, where $\mathds{1}$ is the identity operator.
The probability that message $m'$ is decoded when message $m$ is transmitted is $\mathsf{P}_{m'|m}=\Tr \Pi_{m'} \bm{S}_{\bm{\myin}_m}$. 
The probability of error after sending message $m$ is
\begin{equation*}
\Pem=1-\Tr\left(\Pi_m \bm{S}_{\bm{\myin}_m}\right).
\end{equation*}
We then define $\Pemax$, $\Pemax^{(n)}(R)$ and $E(R)$ precisely as in the classical case. 

With the same reasoning used for classical channels (see \cite{dalai-TIT-2013}) we come to the conclusion that 
\begin{equation}
E(R)\leq \frac{1}{n}\min_{m\neq m'} D_{\text{C}}(\bm{S}_{\bm{x}_m},\bm{S}_{\bm{x}_{m'}})+o(1),
\end{equation}
where $D_{\text{C}}(\cdot, \cdot)$ is now the Chernoff distance between two density operators
\begin{equation}
D_{\text{C}}(A,B)=-\log\inf_{0< s< 1} \Tr A^{1-s}B^s.
\end{equation}
Again we can use bounds on $D_{\text{C}}$ based on additive distances to bound $E(R)$. In particular, we can use the Bhattacharyya distance 
\begin{equation}
\dB(x,x')=-\log \Tr S_x^{1/2} S_{x'}^{1/2},
\end{equation}
for which we have 
\begin{equation}
\dB(x,x')\leq D_C(S_x,S_{x'})\leq 2 \dB(x,x'),
\end{equation}
with equality again on the left for pairwise reversible channels. In the quantum setting, a particularly importance case is given by pure-state channels with states $S_x=|\psi_x\rangle \langle \psi_x|$, for which we always  have $\dB(x,x')=D_C(S_x,S_{x'})=|\langle \psi_x| \psi_{x'}\rangle|^2$. For classical-quantum channels we can also use the bound
\begin{equation}
D_C(S_x,S_{x'})\leq 2 d_{\text{F}}(x,x'),
\end{equation}
where
\begin{eqnarray*}
d_{\text{F}}(A,B) & = & -\log \Tr |\sqrt{A}\sqrt{B}|\\
& = & - \log \Tr \sqrt{\sqrt{A}\,B \sqrt{A}}.
\end{eqnarray*}
We still come to the conclusion, however, that the best choice is simply to use an additive variation of the Chernoff distance 
\begin{equation}
{d}_{\tilde{\text{C}}}(x,x')=D_C(S_x,S_{x'}),
\end{equation}
as for the classical case.


\section{Acknowledgments}
The author would like to thank Telecom Italia Lab and, in particular, Skjalg Leps\o{}y and Gianluca Francini for introducing him to euclidean embedding during the 2008 project DYNAMIC TV. Useful discussions with Richard Blahut and Yury Polyanskiy are also acknowledged.

\appendix[Proof of Lemma \ref{lemma:distances}]

The core part of Lemma \ref{lemma:distances} is by now a classic result
in the theory of positive definite kernels and functions, and should really be interpreted in that context. A detailed discussion can be found for example in \cite{berg-christensen-ressel-book}. Only condition \ref{item3}, which is much important for us, is apparently not usually mentioned in that context. 
In the context of information theory, Jelinek already used the central part of the lemma in his paper \cite{jelinek-1968}. Since we need in any case to add some integration to those references, we provide a complete self contained proof for the reader convenience.

\begin{IEEEproof}[Proof of Lemma \ref{lemma:distances}]
We break down the proof into single implications which, altogether, imply the lemma.

$\bullet\quad$ Implication \ref{item1}$\Rightarrow$\ref{item2}.
\\This is a known connection between infinitely divisible and negative almost definite kernels. Assume $G(\rho)$ is positive semidefinite for all $\rho>0$ and that $\sum_x c(x)=0$. Then
\begin{align*}
\sum_{x,x'}d(x,x')c(x)c(x') 
& = \lim_{\rho\to\infty}\sum_{x,x'}\rho\left(1-e^{-d(x,x')/\rho}\right)c(x)c(x')\\
& = -\rho\sum_{x,x'}e^{-d(x,x')/\rho}c(x)c(x')\\
&\leq 0,
\end{align*}
where we have used the fact that $\sum_{x,x'}c(x)c(x')=0$.

$\bullet\quad$ Implications \ref{item2}$\Leftrightarrow$\ref{item3}.\\
We introduce some notation here and prove a slightly stronger result which will also be useful in the next step of the proof.
For any two functions $q_1$ and $q_2$ on $\mathcal{X}$, let 
\begin{equation}
\label{eq:f_concave}
f(q_1,q_2):=\sum_{x,x'}q_1(x)q_2(x')d(x,x'), \quad f(q):=f(q,q).
\end{equation}

We prove that $f(q)$ is concave on every affine hyperplane defined by $\sum_x q(x)=t$, with $t$ a constant, if and only if $f(q)\leq 0$ whenever $\sum_x q(x)= 0$. Observe that we need only prove midpoint concavity here (a general proof is not substantially different, but requires a more complicated notation).
Note that for two functions $a$ and $b$ on $\mathcal{X}$,  due to the symmetry of $d$ we have 
\begin{equation}
f(a+b)=f(a)+f(b)+2f(a,b)
\end{equation}
and, hence
\begin{equation}
\frac{f(a+b)+f(a-b)}{2}=f(a)+f(b)
\end{equation}

Now, assume $\sum_x q_1(x)=\sum_x q_2(x)=t$, and define the functions $q=(q_1+q_2)/2$ and $c=(q_1-q_2)/2$. Note that $\sum_x q(x)=t$ and $\sum_x c(x)=0$. Then,
\begin{align*}
\frac{f(q_1)+f(q_2)}{2} & =\frac{f(q+c)+f(q-c)}{2} \\
& = f(q)+f(c)\\
&= f\left(\frac{q_1+q_2}{2}\right)+f(c).
\end{align*}
So, if $f(c)\leq 0$ then $f$ is midpoint concave (and hence concave) on the affine hyperplane defined by $\sum_x q(x)=t$. 
Since we used a one to one map $(q_1,q_2)\leftrightarrow(q,c)$, we can invert the reasoning and find, for any $c$, an appropriate pair $q_1$ and $q_2$ to show that if $f$ is concave in any such hyperplane, then $f(c)\leq 0$  whenever $\sum_x c(x)=0$.
%

$\bullet\quad$ Implication \ref{item2}$\Rightarrow$\ref{item4}\\
Define 
\begin{equation}
s(x):=\frac{1}{|\mathcal{X}|}\sum_{x'}d(x,x'),\quad s:=\frac{1}{|\mathcal{X}|^2}\sum_{x,x'}d(x,x'),
\end{equation}
then
\begin{equation}
\tilde{d}(x,x'):=-d(x,x')+s(x)+s(x')-s
\label{eq:d-dtil}
\end{equation}
and finally, for functions $a$ and $b$ on $\mathcal{X}$,
\begin{equation}
\tilde{f}(a,b):=\sum_{x,x'}a(x)b(x')\tilde{d}(x,x'), \quad \tilde{f}(a):=f(a,a).
\end{equation}
Again, by symmetry of $\tilde{d}(x,x')$ we have
\begin{equation}
\tilde{f}(a+b)=\tilde{f}(a)+2\tilde{f}(a,b)+\tilde{f}(b).
\end{equation}
Furthermore, it is not difficult to see that if $\sum_x c(x)=0$, then
$\tilde{f}(c)=-f(c)$, where $f$ is defined as in \eqref{eq:f_concave}. In addition, a direct calculation shows that if $b$ is constant, then
$\tilde{f}(a,b)=0$ for any $a$, which implies that $\tilde{f}(a+b)=\tilde{f}(a)+2\tilde{f}(a,b)+\tilde{f}(b)=\tilde{f}(a)$.

Now, for any $a$, choose $b(x)=\sum_{x'}a(x')/|\mathcal{X}|$, so that $b$ is constant and $\sum_x(a(x)-b(x))=0$. Then, using the properties mentioned above we have
\begin{align}
\tilde{f}(a) & = \tilde{f}(a-b)\\
& = -f(a-b)\\
& \geq 0,
\end{align}
where in the last step we have used the condition \ref{item2} of the Lemma with the choice $c(x)=a(x)-b(x)$.

So, the matrix $\tilde{D}$ with elements $\tilde{d}(x,x')$ is positive semidefinite and, hence, it is a Gram matrix, which means that there exists a set of vectors $\{v_x\}_{x\in \mathcal{X}}$ such that $\tilde{d}(x,x')=\braket{v_x}{v_{x'}}$.
Then, we have
\begin{align}
\|v_x-v_{x'}\|^2 & = \braket{v_x}{v_{x}}-2\braket{v_x}{v_{x'}}+\braket{v_{x'}}{v_{x'}}\\
& = \tilde{d}(x,x)-2\tilde{d}(x,x')+\tilde{d}(x',x')\\
& = -d(x,x)-d(x',x')+2d(x,x')\\
& = 2d(x,x'),
\end{align}
where we have used equation \eqref{eq:d-dtil} and the fact that $d(x,x)=0$ for all $x$ by assumption.
Hence, $d(x,x')=\|u_x-u_{x'}\|^2$ if we set $u_x=v_x/\sqrt{2}$, that is, $d$ is a squared euclidean distance.

$\bullet\quad$ Implication \ref{item4}$\Rightarrow$\ref{item1}\\
This is by now a well known basic property extensively used in the theory of reproducing kernel Hilbert spaces.
We need only prove the implication for $\rho=1$, since $d(x,x')/\rho$ is a squared euclidean distance whenever $d(x,x')$ is. Then, for any $a(x)$ we have
\begin{align}
\sum_{x,x'}a(x)a(x') e^{-d(x,x')}& = \sum_{x,x'}a(x)a(x') e^{-\|v_x-v_{x'}\|^2}\\
& =\sum_{x,x'}\frac{a(x)}{e^{\|v_x\|^2}}\frac{a(x')}{e^{\|v_{x'}\|^2}}e^{2\braket{v_x}{v_{x'}}}\\
& = \sum_{x,x'}b(x)b(x')\sum_{k=0}^{\infty}\frac{(2\braket{v_x}{v_{x'}})^k}{k!}
\end{align}
where $b(x)=a(x)/\exp(\|v_x\|^2)$. Denoting with $\cdot^\otimes$ the $k$-fold Kronecker power of a vector, we can then rewrite the last expression to get
\begin{align*}
\sum_{x,x'}a(x)a(x') e^{-d(x,x')} & =
\sum_{k=0}^\infty  \frac{2^k}{k!}\sum_{x,x'}\braket{(b(x){v_x^{\otimes k}})}{(b(x'){v_{x'}^{\otimes k}})})\\
& = \sum_{k=0}^\infty  \frac{2^k}{k!}\left\|\sum_x b(x) v_x^{\otimes k}\right\|^2\\
& \geq 0.
\end{align*}
Hence the matrix with elements $e^{-d(x,x')}$ is positive semidefinite, and this concludes the proof of the lemma.
\end{IEEEproof}

%


\end{document}